%% file: main.tex
\documentclass[%
reprint,
amsmath,amssymb,
aps,
floatfix,
]{revtex4-2}

\usepackage{graphicx}% Include figure files
\usepackage{dcolumn}% Align table columns on decimal point
\usepackage{bm}% bold math
\begin{document}

\preprint{APS/123-QED}

\title{Connecting the Branches of Multistable Non-Euclidean Origami by Crease Stretching} 

\author{Clark C. Addis}
\author{Salvador Rojas}
\author{Andres F. Arrieta}%
 \email{aarrieta@purdue.edu}
\affiliation{%
Programmable Structures Lab, School of Mechanical Engineering, Purdue University, West Lafayette, IN 47907
}%

\date{\today}

\begin{abstract}
Non-Euclidean origami is a promising technique for designing multistable deployable structures folded from nonplanar developable surfaces. The impossibility of flat foldability inherent to non-Euclidean origami results in two disconnected solution branches each with the same angular deficiency but opposite handedness. We show that these regions can be connected via ``crease stretching'' wherein the creases exhibit extensibility  in addition to torsional stiffness. We further reveal that crease stretching acts as an energy storage method capable of passive deployment and control. Specifically, we show that in a Miura-Ori system with a single stretchable crease, this is achieved via two unique, easy to realize, equilibrium folding pathways for a certain wide set of parameters. In particular, we demonstrate that this connection mostly preserves the stable states of the non-Euclidean system, while resulting a third stable state enabled only by the interaction of crease torsion and stretching. Finally, we show that this simplified model can be used as an efficient and robust tool for inverse design of multistable origami based on closed-form predictions that yield the system parameters required to attain multiple, desired stable shapes. This facilitates the implementation of multistable origami for applications in architecture materials, robotics, and deployable structures.
\end{abstract}

\maketitle

\section{Introduction}

Origami is the ancient Japanese art of paper folding. The simple algebraic kinematics of origami-inspired systems gives rise to its characteristic reconfigurability, which has resulted in numerous applications, in robotic locomotion \cite{melancon2022inflatable,bhovad2021physical,hawkes2010programmable,felton2014method}, metamaterial architecture \cite{schenk2013geometry}, compact deployable structures such as bridges and stadium covers \cite{filipov2015origami}, reconfigurable wheels \cite{lee2017origami}, and -- notably -- simple yet robust inverse shape design algorithms \cite{dudte2021additive, walker2022algorithmic,zhao2022constructing,xiao2022inverse}. Classical origami assumes that shape change stems from folding of an infinitely thin surface involving no deformation or change in strain energy \cite{santangelo2020theory}. Consequently, engineering systems inspired by origami formalism theoretically behave as mechanisms, which by definition do not resist displacement along their degrees of freedom. However, physical embodiments of origami \cite{callens2018flat} do in fact show resistance to displacement, either due to torsional stiffness \cite{brunck2016elastic}, or active elements used to achieve desired kinematic configurations \cite{peraza2014origami}. The former typically shows a single stable shape, while the latter can achieve several configurations at the cost of complex actuation systems and control.

Engineers have investigated the potential of origami with multiple energetic minima, or multistable origami, to add load bearing capacity and pathway configurablity without the need for continuous actuation and complex control. One approach takes advantage of hidden degrees of freedom due to facet bending \cite{silverberg2015origami,liu2018transformation,liu2019invariant,baek2020ladybird}. However, this excludes the utilization of functional materials such as semiconductors (e.g. photovoltaics or transistors) that cannot cope with the large facet strains experienced during bending.

A second approach assumes rigid facets instead relying on elastic bending moments via crease torsion to produce multistability \cite{hanna2014waterbomb,sargent2020origami,fang2017dynamics}. Waitukaitis et al. have shown that crease torsion alone can achieve up to five stable configurations for a 4-facet origami system \cite{waitukaitis2015origami}. However, two key drawbacks remain: (1) the folding paths have multiple solutions which overlap at the flat state \cite{chen2018branches}, and (2) inverse design is difficult due to the coexisting solutions (pathway degeneracy) and the complex multidimensional problem of fine tuning the crease stiffness to match a desired state \cite{waitukaitis2016origami,stern2018shaping,tachi2017self}.

Non-Euclidean origami shows promise to address some of these challenges. This class of origami is still folded from developable, zero Gaussian curvature sheets, but instead of subtending an angle of 2$\pi$ about a particular vertex, as is the case for traditional origami, it is folded from some angular deficiency of less than $2\pi$ (a cone \cite{berry2020topological,waitukaitis2020non}) or some angular excess of greater than $2\pi$ (an E-cone \cite{seffen2016fundamental}). Non-Euclidean origami circumvents the pathway degeneracy of traditional origami because it will always result in two separate folding regimes when folded from a cone no matter how many creases are added \cite{berry2020topological, mcinerney2020hidden}. Intuitively, this is because the two regions represent two mirrored cones [Figure \ref{fig:two-paths} (a)]. The kinematics of non-Euclidean origami preserve the ability to be described by simple algebraic equations \cite{foschi2022explicit,liu2022triclinic, liu2021bio}, and in many instances has suceeded in passively prevent misfolding \cite{liu2018topological, waitukaitis2020non,huang2022design}. This comes at the cost of effectively halving the possible design space.

Resilin, a multipurpose biopolymer commonly found in insect wings \cite{haas1994geometry,rajabi2017dragonfly,mountcastle2014biomechanical, gorb1999serial} serves as an inspiration to expand this design space while enabling multistability. Resilin assigns mountain or valley folds to origami-like insect wings \cite{song2020asian, haas2000function} and serves as an energy storage device, for example in the Rove Beetle \cite{saito2014asymmetric} and the Earwig \cite{haas2000elastic}. However, the Earwig wing is unique in that the energy is stored via stretching in the creases in addition to crease torsion. This crease stretching or ``spring origami'' has been shown to explain the bistability and curved creases displayed by the Earwig wing \cite{faber2018bioinspired}, opening up a vast design space for origami-based systems which localizing all flexibility to the folds.

In this paper, we show that crease stretching can connect the oppositely handed folding regimes of non-Euclidean origami without relying on facet deformation, while retaining the benefit of simple pathway control and straightforward inverse design. This is achieved by establishing a tractable analytical method applied to the well known Miura-Ori unit. We begin with the Miura-Ori folded from the two possible non-Euclidean cones, then relax the rigid crease assumption of classical origami by allowing them to stretch while maintaining infinite facet rigidity. By folding the manifold representing the kinematic space into the energetic space, we can show that inherent symmetries directly predict the existence of a minimum of two, and usually three stable configurations. Furthermore, through derivation of the folding paths, our model allows us to show that crease stretching does not add degeneracy. The analytically predicted configurations and folding paths accurately match benchtop experiments. Our results show that the stable states from the purely non-Euclidean approach are mostly preserved, while adding a third.  Finally, we show how the model is simple enough to be used in inverse design calculations, yet robust enough to be confirmed by experiments.

\section{Alternative Models for Crease Stretching}

We begin with a discussion of alternative models of flexible creases in origami. For sufficiently thin sheets, a straightforward approach is to neglect the effect of stretching, and to assume that the energetics of  crease bending are dominated by torsional deformations. Models based on symmetric Elastica curves make use of rotational springs at curve intersection \cite{jules2019local,jules2020plasticity,dharmadasa2021modeling,gori2022deployment}, while others ignore the bulk and focus on a single nonlinear \cite{iniguez2022rigid,huang2022bio} or linear \cite{lee2022robust,zhang2021generalized,feng2022simplified,zhao2023deployment, walker2018shape} spring whose constants are based on material models. However, we cannot discount crease extensibility during bending because (1) we employ thick creases in this work as thin creases are difficult to manufacture using fused deposition modelling (FDM) 3D printing and (2) we know that kinematically the angular material must be able to stretch to accommodate the transition between non-Euclidean cones. A second approach is to create beam-like models which has been achieved using finite element methods \cite{kim2019bioinspired, rajabi2022insect, daynes2014bio}, or hyperelastic material models \cite{mintchev2018bioinspired}. While highly descriptive and accurate, these beam-like models are computationally expensive and difficult to perform inverse on.

The F-cone model which entails a cone with at least one fold in it \cite{lechenault2015generic} is the closest analytical model that can account for both torsion and stretching. Notably, F-cones remains bistable, even when part of the central fold is removed \cite{yu2022bistability}, a fact that we leverage in our experiments. Nevertheless, the deformation process is still rather complex to model, requiring numerical approximations \cite{walker2020mechanics}, which makes this approach cumbersome for inverse design.

``Spring origami'' offers a computationally simple yet powerful model where the crease is represented by a combination of a rotational and extensional spring \cite{faber2018bioinspired}. Rojas et al. have used it to model 3D printable multistable grippers \cite{rojas2019actuation} and the reconfiguration of temperature sensitive shape memory polymers \cite{rojas2022multistable}. Due to its ability to capture both the torsion and stretching of a system with the minimum possible complexity, we choose to use ``spring origami'' throughout this work.

\section{Problem Definition}

\begin{figure}[b]
\centering
\includegraphics[width=8.6cm]{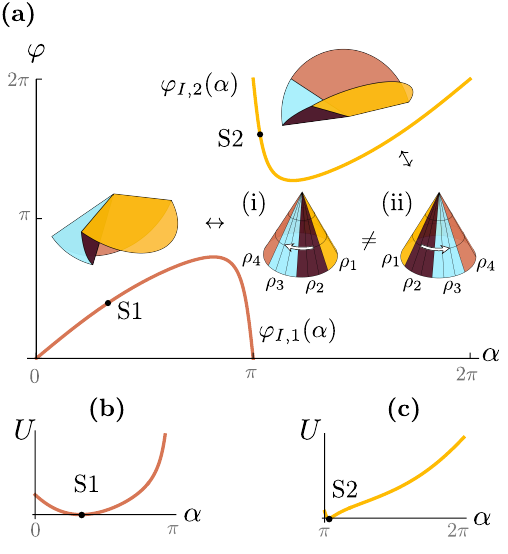}
\caption{\label{fig:two-paths} (a): Non-Euclidean origami with an angular deficiency has two disconnected pathways, each represented by an oppositely handed cone. When a rotational spring is added, two disconnected stable states form (b): The energy landscape for path/cone 1 and stable state S1 (c): The energy landscape for path/cone 2 and state S2.}
\end{figure}

Non-Euclidean origami refers to a 2D surface with sector angles $\rho_i$ subtending an angle of $2\pi + \epsilon$ about the vertex \cite{waitukaitis2020non}, where $\epsilon$ is the angular excess. Equivalently, the surface can be folded from sectors with a subtending angle of $2\pi -\beta$ \cite{faber2018bioinspired}, where $\beta$ is the angular deficit. In general, $\beta = -\epsilon$, and we will use $\beta$ throughout, since pathway disconnection occurs only in systems with angular deficit \cite{berry2020topological}. We choose to examine a Miura-Ori unit, an origami fold pattern originally developed for folding membranes in space mechanisms \cite{koryo1985method}, given the extensive attention it has received in the literature and its universal applications. For a symmetric Miura-Ori unit, one additional geometric parameter $\gamma$  \cite{schenk2013geometry} is required to define the symmetric facets $\rho_1 = \rho_4$ [shown in Figure \ref{fig:two-paths} (a) (i)]. Using spherical trigonometry for the basic case of a non-Euclidean system without crease stretching we can write the kinematics (see derivation details in SI A.3.5) as:

\begin{eqnarray}
\label{eq:intersect-path}
\varphi_{I,n}(\alpha) &=& 2 \tan ^{-1}
    \Bigg[ 
    \frac{\sin \alpha}{-\cot(\gamma + \beta/2) \sin \gamma - \cos\gamma \cos\alpha}
    \Bigg] +
    \nonumber\\
 &&  2\pi(n-1),
\end{eqnarray}

with $n=1$ or $n=2$, where $\varphi$ and $\alpha$ are the dihedral angles defined in Figure \ref{fig:space-folding} (a). Notice, that Eq. (\ref{eq:intersect-path}) results in two folding pathways: $\varphi_{I,1}(\alpha)$ (left) and $\varphi_{I,2}(\alpha)$ (right), plotted in Figure \ref{fig:two-paths}a\ (a) as Eq. (\ref{eq:intersect-path}) with $\gamma = 3\pi/4$, and $\beta = 10^{\circ}$. This matches prior work establishing that for $\beta > 0$ two disconnected pathways always exist \cite{berry2020topological}. The disconnected pathways can be abstracted as two ``oppositely handed'' cones. For path 1, when viewed with the vertex at the top, $\rho_i$ cyclically increases in the clockwise direction around the directrix [Figure \ref{fig:two-paths} (a) (i)], whereas for path 2, $\rho_i$ cyclically increases in the counterclockwise direction around the directrix [Figure \ref{fig:two-paths} (a) (ii)].

We consider the addition of a rotational spring between facets $\rho_2$ and $\rho_3$, with an equilibrium angle $\theta_0 = 79.76^{\circ}$ which reveals two stable states S1 and S2. The equilibrium angle $\alpha_{0,1}$ for stable state S1 on pathway 1 can be calculated using:

\begin{eqnarray}
\label{eq:alpha-0-1}
\alpha_{0,1} &=&
    \cos^{-1}
    \left[
    \frac{\cot\gamma\sin\eta}
    {\sqrt{\cos^2\eta + \cot^2(\theta_0/2)}}
    \right] - 
    \nonumber\\
    &&\tan^{-1}\left[\frac{\cot(\theta_0/2)}{\cos\eta}\right],
\end{eqnarray}

and the equilibrium angle $\alpha_{0,2}$ for stable state S2 on pathway 2 can be calculated as:

\begin{eqnarray}
\label{eq:alpha-0-2}
\alpha_{0,2} &=&
    2\pi - \cos^{-1}
    \left[
    \frac{\cot\gamma\sin\eta}
    {\sqrt{\cos^2\eta + \cot^2(\theta_0/2)}}
    \right] - 
    \nonumber\\
    &&\tan^{-1}\left[\frac{\cot(\theta_0/2)}{\cos\eta}\right].
\end{eqnarray}

Eqs. (\ref{eq:alpha-0-1}) and (\ref{eq:alpha-0-2}) are derived by using spherical trigonometry and the harmonic identity to locate the coordinates where the gradient of the energy is zero as detailed in SI B.3. Notice, however, that in Figures \ref{fig:two-paths} (b) and (c) the two stable states lie on two different pathways. Thus, these two stable states could never be physically realized on the same rigid system. These disconnected pathways (i.e., disjointed sets) are an intrinsic characteristic of n-fold, non-Euclidean origami with an angular deficiency \cite{berry2020topological}. In the following, we establish a method to connect these disjointed regions in a physical system while retaining bistability by allowing crease stretching.

\section{Parameter Space}

\begin{figure}[b]
\includegraphics[width=8.6cm]{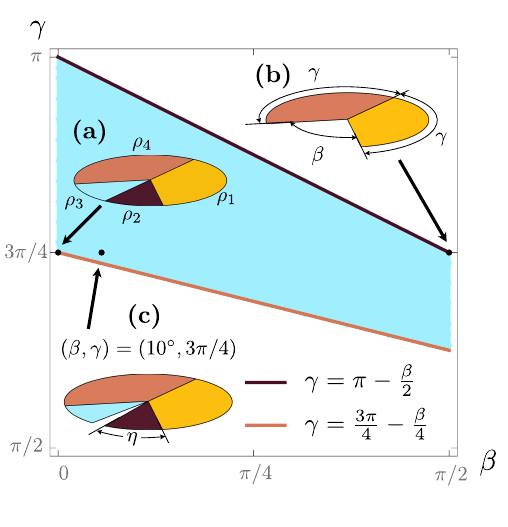}
\caption{\label{fig:params} The valid parameter space for this analysis is highlighted by the shaded area. There are two boundaries: the upper boundary  corresponds to $\eta = 0$, and the lower boundary simplifies the kinematic analysis. The point at $\beta = 10 ^{\circ}$ and $\gamma = \frac{3\pi}{4}$ is used throughout this study.}
\end{figure}

We first define the derived parameter $\eta$, which is equivalently the facet angle subtending $\rho_2$, and the facet angle subtending $\rho_3$ [Figure \ref{fig:params} (a)]. We define that in the flat state, $\eta$, $\gamma$, and $\beta$ lie in a plane [see Figure \ref{fig:params} (b) for definition of $\gamma$ and $\beta$ and Figure \ref{fig:params} (c) for definition of $\eta$]. Thus, they are explicitly related by:

\begin{equation}
    \label{eq:eta}
    2\eta + 2\gamma + \beta = 2\pi.
\end{equation}

We introduce parametric bounds on our system, by first enforcing that $\eta > 0$, implying that facets $\rho_2$ and $\rho_3$ will always exist. Using Eq. (\ref{eq:eta}), this definition is equivalent to the statement: 

\begin{equation}
    \gamma < \pi - \beta/2.
    \label{eq:non-zero-eta}
\end{equation}

To allow for the space folding technique employed in section \ref{sec:space-folding}, we impose that the kinematic space always form a closed region using the two kinematic boundaries we establish in section \ref{sec:boundaries} (see SI A.4.1 for details) resulting in the additional restriction that:

\begin{equation}
    \gamma > 3\pi/4 - \beta/4.
    \label{eq:simplicity}
\end{equation}

Finally, to further simplify our analysis (see SI A.3.3 for full justification), we restrict $\beta$ and $\gamma$ to the domains: 

\begin{equation}
    0 < \beta < \pi/2 \quad \mathrm{and} \quad \pi/2 <\gamma <\pi.
    \label{eq:beta-gamma-res}
\end{equation}

The four constraints given by Eqs. (\ref{eq:non-zero-eta}-\ref{eq:beta-gamma-res}) form a closed region represented by the shaded area in Figure \ref{fig:params}. The geometric parameters used throughout this article are $\beta = 10^{\circ}$ and $\gamma = 3\pi/4$ [Figure \ref{fig:params} (c)].

\section{Kinematics and Energetics}\label{sec:boundaries}

\begin{figure*}
\includegraphics{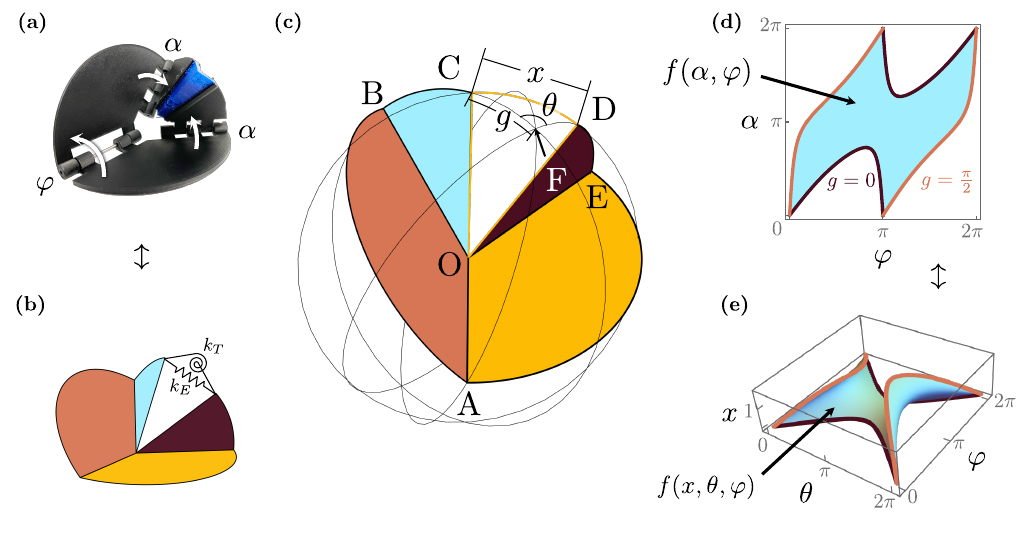}% Here is how to import EPS art
\caption{\label{fig:space-folding} (a): We treat the stretchable crease as a torsional and extensional spring with stiffness $k_T$ and $k_E$ (b): The 2 degrees of freedom, $\varphi$ and $\alpha$ completely define the kinematic space (c): If all of the dihedral edges are unit length, all endpoints lie on a sphere. The spring $k_E$ acts over distance $x$, which is the chord from point C to point D, and the spring $k_T$ acts about angle $\theta$ which is the angle defined by $\triangle CFD$ about point $F$. The parameter $g$ is useful in establishing the kinematic bounds, and it is equal to arc $FC=FD$ (d): Not all $\varphi-\alpha$ pairs are valid. When $g=0$ the facets intersect, and when $g=\pi/2$ the extensional spring clips through the facets, both of which are physically impossible (e): We can fold the $\alpha-\varphi$ kinematic space into $x-\theta-\varphi$ space to simplify the energetic analysis.} 
\end{figure*}

The kinematic analysis of a stretchable crease unit requires the definition of two independent degrees of freedom $\varphi$ and $\alpha$ [Figure \ref{fig:space-folding}(a)]. This effectively creates a cut at $\rho_2$ and $\rho_3$. Allowing $\varphi$ and $\alpha$ to both be members of the open set $(0,2\pi)$ provides all possible configurations of the system. 

To represent a flexible crease [the material between $\rho_2$ and $\rho_3$ in Figure \ref{fig:space-folding} (a)], we use a rotational spring of stiffness $k_T$ with equilibrium angle $\theta_0$ and an extensional spring of stiffness $k_E$ with equilibrium distance $x_0$ [Figure \ref{fig:space-folding} (b)], which act on $\theta$ and $x$ respectively [Figure \ref{fig:space-folding} (c)]. We define $\theta$ to be the angle subtended by the spherical arc $CD$ from point F, while we define $x$ as the Euclidean distance between points C and D. Finally, we assume the restoring force from the stretchable crease to be of a significantly higher order of magnitude than the potential due to gravity allowing us to ignore  the effect of mass in the facets ($\rho_i$). The resulting energetics are captured by:

\begin{equation}
    \label{eq:energy-text}
    U = \frac{1}{2}\left(k_E[x(\alpha,\varphi)-x_0]^2 + k_T[\theta(\alpha,\varphi)-\theta_0]^2\right).
\end{equation}

Interestingly, assuming rigid facets and lumped springs impose two kinematic constraints on the system. These two constraints can be described by the parameter $g=FC=FD$ [ Figure \ref{fig:space-folding} (c)]. The first constraint is that facets $\rho_2$ and $\rho_3$ cannot intersect, corresponding to $x=0$, or $g=0$ in Figure \ref{fig:space-folding} (d). Equivalently, this boundary is the kinematic space of a non-Euclidean system without crease stretching, and thus is directly defined by Eq. (\ref{eq:intersect-path}). We refer to this as the ``intersect boundary,'' $\varphi_I$. The second constraint is $g=\pi/2$, which corresponds to when facets $\rho_2$ and $\rho_3$ are parallel to each other, rendering impossible the placement of a rotational spring. We refer to this as $\varphi_S$, for ``spring boundary,'' and it is defined by (see SI A.3.4 for this derivation):

\begin{equation}
    \label{eq:phi-spring}
    \varphi_{S} (\alpha) = 
    \pi - 2 \tan ^{-1}
    \Bigg( 
    \frac{\tan(\gamma + \beta/2) \sin \gamma - \cos\gamma \cos\alpha}{\sin \alpha}
    \Bigg).
\end{equation}

We plot Eqs. (\ref{eq:intersect-path}) and (\ref{eq:phi-spring}) in $\alpha-\varphi$ space [Figure \ref{fig:space-folding} (d)] and show that they always form a closed region, given the parameter limits outlined in Figure \ref{fig:params} (see SI A.4 for details).

\section{Stability Analysis}\label{sec:space-folding}

\begin{figure}[b]
\centering
\includegraphics[width=8.6cm]{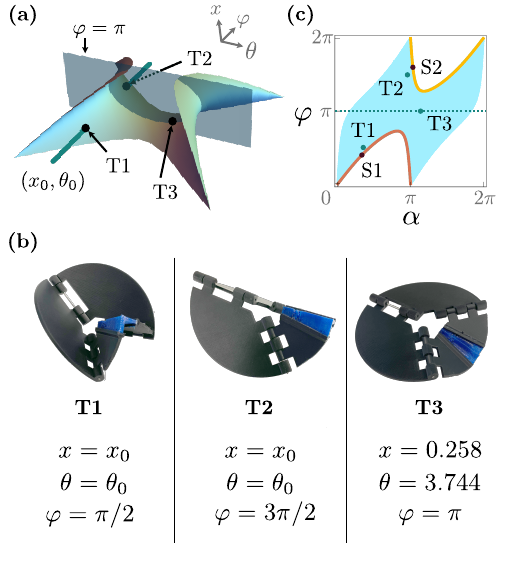}
\caption{\label{fig:stability} (a): Our model predicts 3 stable states, two of which result from the global minimum (T1, T2) and one which is a local minimum of the balance of crease torsion and stretching (T3). (b): We can see visually that the global minimum of $x_0$ and $\theta_0$ occurs twice (T1, T2). We can show mathematically that the global minimum can only occur along the plane $\varphi = \pi$ (T3). (c): This model is verified by experiments.}
\end{figure}

Transforming the kinematics from $\alpha-\varphi$ to $x-\theta$ coordinates simplifies the stability analysis because the energetics of the problem, as defined in Eq. (\ref{eq:energy-text}), are in  $x$ and $\theta$. However, $x-\theta$ coordinates do not provide a one-to-one mapping to $\alpha-\varphi$ coordinates as described in SI A.5. This motivates our choices to use $x-\theta-\varphi$ coordinates to represent the system, which effectively folds $\alpha-\varphi$ space into 3 dimensions, as shown in Figure \ref{fig:space-folding} (e), and provides a one-to-one mapping to $\alpha-\varphi$ space. To characterize this surface, we can write a level curve $f(x,\theta,\varphi) = 0$, using the spherical law of sines, as detailed in SI A.7, yielding:

\begin{equation}
    f(x,\theta,\varphi) = \frac{\sin(\theta/2)}{\sin\gamma} - \frac{\sin(\varphi/2)}{\sin\left(\eta + \sin^{-1}\left[\frac{x}{2\sin(\theta/2)}\right]\right)} = 0.
    \label{eq:f-of-x-theta-phi}
\end{equation}

Prior work has established that energy minima, and thus stable states, occur when the gradient of energy ($\nabla U$) and the gradient of the kinematic space ($\nabla f$) point in the same direction \cite{li2020theory}. Using the dot product, this statement can be equivalently written as:

\begin{equation}
    \label{eq:dot-product}
    \nabla U \cdot \nabla f = 
    \left \lvert \nabla U\right \lvert 
    \left \lvert \nabla f \right \lvert,
\end{equation}

where the gradient operator is defined in $x-\theta-\varphi$ coordinates. Applying this definition of the gradient to Eq. (\ref{eq:energy-text}) and Eq. (\ref{eq:f-of-x-theta-phi}), yields $\nabla f$ and $\nabla U$ as:

\begin{equation}
    \label{eq:nabla-U-text}
    \nabla U = \left[k_E(x-x_0),k_T(\theta-\theta_0),0\right]
\end{equation}

and

\begin{equation}
    \label{eq:nabla-f-text}
    \nabla f = \left[\frac{\partial f}{\partial x}, \frac{\partial f}{\partial \theta}, \frac{\partial f}{\partial \varphi}\right].
\end{equation}

Notice that Eq. (\ref{eq:nabla-U-text}) establishes that the energy gradient is always a 2D vector, and Eq. (\ref{eq:nabla-f-text}) reveals that kinematic gradient is in general a 3D vector. Therefore, the two vectors point in the same direction only when the kinematic vector degenerates into a 2D vector in the $x-\theta$ plane. Inspection of Eq. (\ref{eq:nabla-f-text}) reveals that this collapse occurs when $\partial f/\partial\varphi = 0$. Our analysis in SI B.1.2 shows that $\partial f/\partial\varphi = 0$ occurs when $\varphi=\pi$. We can visually establish in Figure \ref{fig:stability} (a) that this stable state (T3) indeed occurs within the plane $\varphi =\pi$ in $x-\theta-\varphi$ space, and in Figure \ref{fig:stability} (b) on the line $\varphi =\pi$ in $\alpha-\varphi$ space.

To uncover the other stable states, we re-examine Eq. (\ref{eq:dot-product}) and determine that $\nabla f =0$ or $\nabla U =0$ also satisfy this expression. We show in SI B.1.3 that $\nabla f$ is never zero and from inspection of Eq. (\ref{eq:nabla-U-text}) we observe that $\nabla U =0$ when $x=x_0$ and $\theta =\theta_0$. This result implies that additional stables states appear according to the number of $(\alpha,\varphi)$ pairs that can produce a specific $(x,\theta)$ pair. Specifically, we reveal in SI A.5 that every $(x,\theta)$ pair, corresponds to two $(\alpha,\varphi)$ pairs, with the notable exception at $\varphi =\pi$, i.e., the flat folded configuration. Figure \ref{fig:stability} (a) reveals a visual intuition for why this is true: the solid line, which represents a given $(x_0,\theta_0)$, intersects the surface $f$ twice, at T1 and T2 and is symmetric about $\varphi =\pi$. The equations used to obtain $(\alpha_1, \varphi_1)$ and $(\alpha_2, \varphi_2)$ for T1 and T2 respectively, are derived using Napier's analogies and the spherical law of sines in SI A.5, which yields the following conditions:

\begin{equation}
\varphi_1
= 
2\sin^{-1}
\left[
    \frac{\sin\frac{\theta_0}{2}}{\sin\gamma}
    \sin
    \left(
        \eta
        + \sin^{-1}
        \left[
            \frac{x_0}{2\sin\frac{\theta_0}{2}} 
        \right]
    \right)
\right],\
\label{eq:phi-1}
\end{equation}
and
\begin{equation}
\varphi_2
= 
2\pi - \varphi_1,
\label{eq:phi-2}
\end{equation}
with
\begin{equation}
    \alpha_{1,2} = 2 \tan^{-1}\left[
        -\frac{
        \cos\left(\frac{1}{2}[\gamma \pm \eta]\right)
    }{
        \cos\left(\frac{1}{2}[\gamma \mp \eta]\right)
    } \tan\left(\frac{\varphi_1}{4} - \frac{\theta_0}{4}\right)\right].
    \label{eq:alpha-1}
\end{equation}

To obtain state T3, we write U as a function of $\alpha$, holding $\varphi$ constant at $\pi$ and setting $\partial U /\partial \alpha =0$ (SI B2). Note that this requires choosing a ratio of $k_E/k_T$. We chose $10^3$ because it is close to the range considered in Ref. \cite{faber2018bioinspired} for which the predictions obtained with our model reveal the parameter values for the three stable states, as summarized in Table \ref{tab:stable-table}.

The predicted parameter values by our closed-form solutions enable the design of the experimental demonstrator shown in Figure \ref{fig:stability} (b). In this demonstrator, we observe the symmetry of states T1 and T2 about $\varphi = \pi$ which have identical $(x, \theta)$ values. Additionally, the third stable state (T3), is found at $\varphi=\pi$. These experimental observations match closely the theoretical model predictions in Table \ref{tab:stable-table}.

\begin{table}[b]%The best place to locate the table environment is directly after its first reference in text
\caption{\label{tab:stable-table}%
Degrees of freedom values and stability conditions predicted by our model for the three stable states for a system with parameters $\gamma =3\pi/4$, $\beta=10^{\circ}$, $x_0 =0.25$, $\theta_0=1.39$, and $k_E/k_T = 10^3$.
}
\begin{ruledtabular}
\begin{tabular}{lrrrrr}
\textrm{State}&
\textrm{$x$}&
\textrm{$\theta$}&
\textrm{$\alpha$}&
\textrm{$\varphi$}&
\textrm{$\nabla U,\nabla f$}
\\
\colrule
T1 & 0.250   & 1.39  & 1.10 & $\pi/2$  & $\nabla U = 0$ \\
T2 & 0.250   & 1.39  & 3.01 & $3\pi/2$ & $\nabla U = 0$ \\
T3 & 0.258  & 3.74  & 3.57 &$\pi$ & $\nabla U/|\nabla U| = \nabla f/|\nabla f|$\\
\end{tabular}
\end{ruledtabular}
\end{table}

We now compare the stable states found by strict non-Euclidean origami [Figure \ref{fig:two-paths} (a)] to the stable states for an equivalent system with the addition of crease stretching [Figure \ref{fig:stability} (c)]. Note that the shaded region in Figure \ref{fig:stability} (c) represents the kinematic space for crease stretching whereas the solid lines represents the kinematic space of strict non-Euclidean origami. Without crease stretching, we observe two disconnected stable states, S1 and S2. In contrast, allowing the crease to stretch connects the states via the shaded region (T1 and T2) and introduces an additional, kinematically accessible, third stable state (T3). Consequently, crease stretching connects the otherwise disjointed two stable states, i.e., S1 and S2.

\section{Folding Pathways}

\begin{figure}
\includegraphics{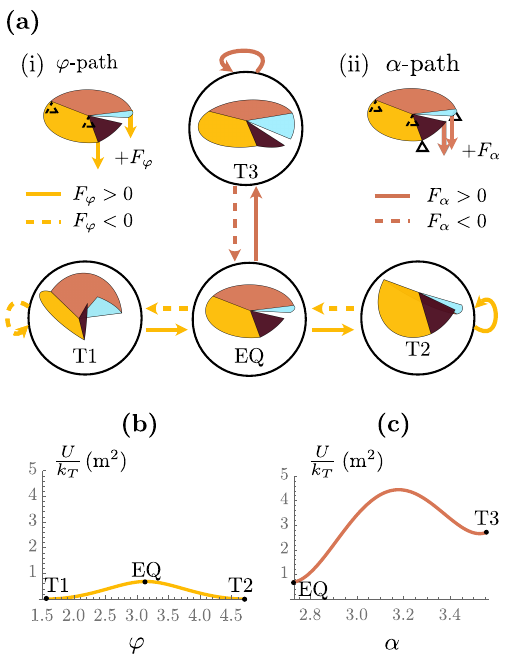}% Here is how to import EPS art
\caption{\label{fig:wide-fsm} (a): Stable states T1, T2, and T3 are connected by two main pathways (i) the $\varphi$-path where $\varphi$ is the only DOF, where when a +$F_{\varphi}$ is applied results in system folding from T1 to EQ and EQ to T2, and (ii) the $\alpha$-path where $\alpha$ is the only DOF, which when a positive $F_{\alpha}$ is applied results in folding from EQ to T3 (b): Energy landscape along the $\varphi$-path, showing local maximum at EQ and local minima at T1 and T2 (c): the energy landscape along the $\alpha$-path, showing local minima at EQ and T3.}
\end{figure}

Figure \ref{fig:stability} (b) showed that the three stable states have no kinematic obstacles between them. Here, we elucidate the two equilibrium folding pathways by which these states are connected using the Principle of Virtual Work (SI C). This approach reveals that there are two possible equilibrium pathways, one for each degree of freedom. One pathway corresponds to $\frac{\partial U}{\partial \alpha} = 0$. A solution to this equation fixes points O and A and applies two vertical downward forces, $F_{\varphi}$, at B and E [Figure \ref{fig:wide-fsm} (a) (i)]. We define this to be the $\varphi$-path because $\varphi$ is our degree of freedom. The energy along this pathway is shown in Figure \ref{fig:wide-fsm} (b). Notice that the endpoints of this path are T1 and T2, the two stable states established earlier. The other path corresponds to $\frac{\partial U}{\partial \varphi} = 0$. This pathway requires fixing points O, A, B, and E, and applying two vertical forces $F_\alpha$ at C and D, see Figure \ref{fig:space-folding} (c) for the definition of points A-E and O. We define this to be the $\alpha$-path [Figure \ref{fig:wide-fsm} (a) (ii)] because $\alpha$ is our degree of freedom. The energy along this pathway is shown in Figure \ref{fig:wide-fsm} (c). The full energy landscape for the entire kinematic space is shown in Figure \ref{fig:design} (a).

Notice that in Figure \ref{fig:wide-fsm} (c), state T3 is not connected directly to states T1 or T2. Instead, its endpoint is an intermediate saddle point EQ. Figure \ref{fig:wide-fsm} (a) summarizes how T1, T2, T3, and EQ are connected by the $\alpha$ path and the $\varphi$ path. For instance, if we desire to move from state T1 to state T2, we would apply a positive $F_{\varphi}$ to the system with the boundary conditions of the $\varphi$-path. Application of a positive $F_{\varphi}$ past state T2, results in deflection back to state S2 upon release (See Video S1). Figure \ref{fig:wide-fsm} (a) also indicates that to access state T3 from T1 requires the application of a positive $F_{\varphi}$ under the boundary conditions of the $\varphi$-path until the system is at the saddle point, EQ. Then, by switching a positive $F_{\alpha}$ with the boundary conditions of the $\alpha$-pathway, the system reaches state T3. Continuing to apply a positive $F_{\alpha}$ and subsequent release results in  self-equilibrating deflection into state T3 (See video S2). For a given transition from one state to another, Figure \ref{fig:wide-fsm} (c) can be read by starting at the initial state, and then applying $\pm F_{\alpha}$ or $\pm F_{\varphi}$ until the desired final state is reached.

\section{Inverse Design}

\begin{figure}[b]
\centering
\includegraphics[width=8.6cm]{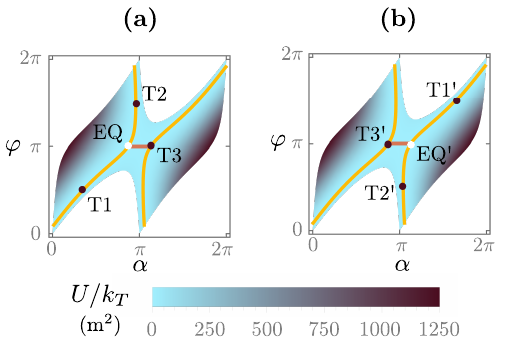}
\caption{\label{fig:design}  Folding pathways, resulting stable states, and normalized strain energy choosing $\varphi_{0,1}$ (value of $\varphi$ at T1)  and $x_0$ as design objectives and $\theta_0$ as our design variable. (a) For the design objective of $\varphi_{0,1} = \pi/2$ and $x_0 = 0.25$ inverse design results in $\theta_0 = 1.39$  (b) The same design objective is also achieved using $\theta_0 = 2\pi -1.39$. By choosing one (a) or (b), we can control whether the two stable states have $\alpha$ values in the open interval (a): $(0,\pi)$ or (b): $(\pi,2\pi)$.}
\end{figure}

We now shift to the ultimate focus of this paper: to inversely design a spring origami Miura-Ori unit. We begin with a justification of our choice of design objective. Since there are a small number of elastomeric materials available for our manufacturing process (FDM 3D printing), we relate $k_E$ and $k_T$ at a ratio of $k_E/k_T = 10^3$, as before. We have also shown that choosing the shape of either T1 or T2 [$(\varphi_{T1},\alpha_{T1})$ or $(\varphi_{T2},\alpha_{T2})$] implies defining the other state's shape given their $2\pi$ symmetry in $\varphi$ [see Eq. (\ref{eq:phi-2}) and Eq. (\ref{eq:alpha-1})]. Therefore, the full range of available design variables are:

\begin{equation}
    \gamma, \beta, x_0, \theta_0, \quad\mathrm{and}\quad (\varphi_{T1} |  \varphi_{T2} | \alpha_{T1} | \alpha_{T2})
    \label{eq:dsn-variables}
\end{equation}

To perform inverse design, we can now choose any four variables from Eq. \ref{eq:dsn-variables} and determine the fifth. We illustrated this process, fixing $\gamma$, $\beta$, $x_0$, and $\varphi_{T1}$ and calculating the required $\theta_0$. To simplify our kinematics, we keep the parameters fixed at $\gamma=3\pi/4$ and $\beta=10^{\circ}$ as in Figure \ref{fig:params} (c). We select $\varphi$ as our design objective over $\alpha$ since all of its facets are connected to standard origami linkages, so linking the units kinematically to an origami string \cite{kamrava2018programmable}, or an origami metamaterial \cite{schenk2013geometry} would be a simple task. We choose $x_0$ as the second design objective, as limitations from the herein employed fabrication method, imposes restrictions on manufacturing larger $x_0$ values. However, freer choice of manufacturing processes would allow for the selection of $\theta_0$ as design objective. To illustrate this process, we explicitly derive the $\theta_0$ required to achieve a given $\varphi_{0,1}$-$x_0$ pair, as detailed in SI A.8, yielding:

\begin{subequations}
\label{eq:inverse-design}
\begin{eqnarray}
\label{eq:inverse-design-low}
\theta_0(x_0,\varphi_{0,1})=&&2 \sin^{-1}\left(\frac{x_0}{2}\frac{\sqrt{A^2 +1}}{A}\right)\quad 
\\
&&\textrm{or}\nonumber\\&&
\label{eq:inverse-design-high}
 2\pi - 2\sin^{-1}\left(\frac{x_0}{2}\frac{\sqrt{A^2 +1}}{A}\right) 
\end{eqnarray}
\end{subequations}

where 

\begin{equation}
A = \frac{\sin\eta}{\frac{2}{x_0}\sin\frac{\varphi_{0,1}}{2}\sin\gamma - \cos\eta}.
\end{equation}

Notice that Eq. (\ref{eq:inverse-design}) reveals there are two possible $\theta_0$ values that satisfy our desired $\varphi_{0,1}$ and $x_0$, where the first applies for $\theta_0\in(0,\pi)$ and the second for $\theta_0\in(\pi,2\pi)$.

To understand the difference between these two values of $\theta_0$, we choose a representative design point of $\varphi_{0,1}=\pi/2$ and $x = 0.25$ for simplicity. Using Eq. (\ref{eq:inverse-design-low}) we obtain the value $\theta_0 = 1.39$, and using Eq. (\ref{eq:inverse-design-high}) we obtain the value $\theta_0 = 2\pi - 1.39$, both of which satisfy our design requirement. Figure \ref{fig:design} (a) shows what the stable states and the folding pathways look like for $\theta_0 = 1.39$. Recall that $\theta_0 = 1.39$ and $x=0.25$ were the parameters used in the preceding kinematic analysis, and that stable states, T1, T2, and T3, were indeed at $\varphi = \pi/2$,  $\varphi = 3\pi/2$ and $\varphi =\pi$, as shown in Figure \ref{fig:stability} (c), showcasing the predictive power of the inverse design. Figure \ref{fig:design} (b) demonstrates the impact of the alternative choice of $\theta_0 =2\pi-1.39$. Notice that T1 and T2 switch are now located on the right-hand path, while T3 now lies on the left-hand path. From this, we can conclude that if $\theta_0 \in(0,\pi)$ [Eq. (\ref{eq:inverse-design-low})] then the two symmetric stable states will have $\alpha\in(0,\pi)$, whereas if we chose the  $\theta_0 \in(\pi,2\pi)$ [Eq. (\ref{eq:inverse-design-high})] we will have that the two symmetric stable states have $\alpha\in(\pi,2\pi)$.

\section{Conclusion}

We show via analysis and experiments a simplified technique of modeling Miura-Ori units with stretchable creases. Our analysis is based on a tractable model enabling the closed-form prediction of the stable state and folding paths of 4-vertex, non-Euclidean origami units with crease stretching. The derived model can accurately predict almost all of the observed behavior of a physical system. The model also enables an inverse design technique that gives two symmetric solutions for a desired shape. In theoretical terms, this is accomplished by connecting the disconnected regimes of non-Euclidean origami via crease stretching. This work provides an analytical model to access the design space of multistable origami, and lays the framework to allow for efficient inverse design of desired shapes based on accessible closed-form solutions the stability of which does not require actuation systems or facet deformation. 

\section{Acknowledgements}
We acknowledge funding from AB-Inbev and the Purdue Winkelman fellowship.

\section{Author Contributions}
A.F.A and C.C.A. conceptualized the research, C.C.A conducted the model derivation, A.F.A. supervised the research, C.C.A and S.R. developed the experimental demonstrator, and A.F.A. and C.C.A prepared the manuscript. 

\newpage

\bibliography{biblo}% Produces the bibliography via BibTeX.

\end{document}

\maketitle

\tableofcontents

\appendix

\section{Kinematics}

\subsection{Parameters and Degrees of Freedom}
\label{section:parms-DOF}

\input{appendicies/kinematics/setup.tex}

\subsection{Spherical Intepreation}

\input{appendicies/kinematics/spheres.tex}

\subsection{Certain $\alpha-\varphi$ pairs are inaccessible}

\input{appendicies/kinematics/bounds.tex}

\subsection{The Spring Boundary and Intersect Boundary Form a Closed Region}
\label{section:closed-region}

\input{appendicies/kinematics/closed-region.tex}

\subsection{$x$ and $\theta$ do not always uniquely define the system}
\label{section:x-theta-symmetry}

\input{appendicies/kinematics/x-theta.tex}

\subsection{Converting $\varphi-\alpha$ to $x-\theta$}

\input{appendicies/kinematics/alpha-phi}

\subsection{$x-\theta-\varphi$ Kinematics}

\input{appendicies/kinematics/x-theta-phi.tex}

\section{Energentics and Stable States}

\subsection{The Stable States are at $\varphi = \pi$ and $(x,\theta) = (x_0, \theta_0)$}

\subsubsection{The Three Conditions For Stable States}

\input{appendicies/energetics/energetics-intro.tex}

\subsubsection{The Energetic Local Minimum ($\nabla f\cdot \nabla U = |\nabla U| |\nabla f|$) is Equivilent to $\varphi = \pi$}\label{section:pi-equiv}

\input{appendicies/energetics/combined-gradient.tex}

\subsubsection{The Kinematic Local Minimum ($\nabla f = 0$) Can be Excluded}

\input{appendicies/energetics/kinematic-gradient.tex}

\subsubsection{Global Minimum ($\nabla U = 0$)}

\input{appendicies/energetics/global.tex}

\subsection{Finding the Local Minimum of Crease Stretching}

\input{appendicies/energetics/ke_kt_dependence.tex}

\subsection{Finding the Local Minimum without Crease Stretching}

\input{appendicies/energetics/sans_stretch.tex}

\section{Virtual Work and Folding Pathways}

\input{appendicies/folding-paths/general}

\bibliography{biblo}{}
\bibliographystyle{plain}

%% file: supp/appendicies/kinematics/setup.tex
\subsubsection{Parameters}

\begin{figure}
    \centering
    \includegraphics{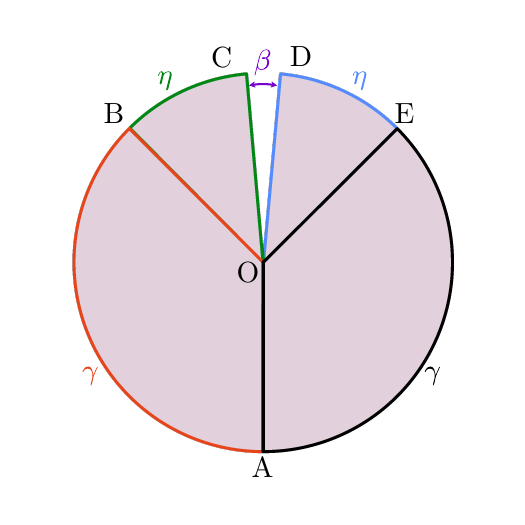}
    \caption{Geometry of the System}
    \label{fig:geometry}
\end{figure}

Figure \ref{fig:geometry} shows the system in it's flat configuration. We define three ``paramters'' for this system, $\gamma$,  $\eta$, and $\beta$. $\gamma$ is defined to be the arclength between OA and OB, $\eta$ is the arclength of BC and DE, and $\beta$ is the angular deficit of the material when folded flat. The relationship between the parameters can be written mathematically as

\begin{equation}
	2\pi = 2\eta +2\gamma + \beta
\end{equation}

or 

\begin{equation}
	\eta = \pi - \gamma - \beta/2
    \label{eq:eta}
\end{equation}

To simplify analysis, we put the following limits on $\beta$ and $\gamma$

\begin{equation}
    \beta \in (0,\pi/2)\quad\gamma \in (\pi/2,\pi)
    \label{eq:base-limits}
\end{equation}

For this model to work, we must enforce that facet OBC always exist, that is that $\eta$ is always some nonzero positive value. Another way of saying this is that $\eta>0$. Enforcing this on EQ \ref{eq:eta}, we get that

\begin{equation}
    \pi - \gamma -\beta/2 > 0 
\end{equation}

or 

\begin{equation}
     \gamma + \beta/2 < \pi
     \label{eq:beta-gamma-pi}
\end{equation}

or 

\begin{equation}
     \gamma < \pi - \frac{\beta}{2}
     \label{eq:upper-gamma}
\end{equation}

we also can easily deduce from EQ \ref{eq:base-limits} that $\gamma +\beta/2 > \pi/2$. To conclude, we can say that

\begin{equation}
     \pi/2 <\gamma + \beta/2 < \pi
     \label{eq:beta-gamma-constraints}
\end{equation}

There is an additional constraint on this system, which is that 

\begin{equation}
    \gamma>\frac{3\pi}{4} - \frac{\beta}{4}
    \label{eq:lower-gamma}
\end{equation}

It is not obvious here why this constraint exists. Here, all we can say is that it simplifies the energetic analysis. The full justification for this is provided in Section \ref{section:lower-limit-justify}

\begin{figure}
    \centering
    \input{supp/figs/tikz/valid-beta-gamma.tex}
    \caption{The shaded region shows the intersection of $\pi/2<\gamma < \pi$, $0<\beta<\pi/2$, $\gamma > 3\pi/4 -\beta/4$, and $\gamma<\pi-\beta/2$, which is the valid parameter space for this model}
    \label{fig:param-limits}
\end{figure}

To summarize, we can take EQ \ref{eq:base-limits}, EQ \ref{eq:lower-gamma}, EQ \ref{eq:upper-gamma} and show there overlap graphically  in Figure \ref{fig:param-limits}. Any $(\beta,\gamma)$ pair which satisfies these constraints will be allowed.

\subsubsection{Degrees of Freedom}

\begin{figure}
    \centering
    \includegraphics{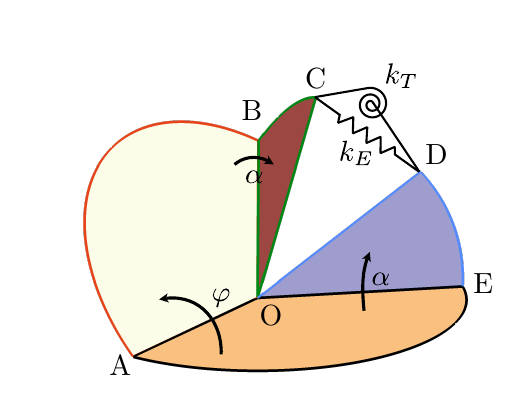}
    \caption{Degrees of Freedom and Springs}
    \label{fig:DOF}
\end{figure}

Distinct from the parameters, are the degrees of freedom (DOF) of the system. They are completely independent of the parameters, and define the kinematic state of a given system. We define two degrees of freedom, $\varphi$ and $\alpha$ (Figure \ref{fig:DOF}). 

$\varphi$ is defined to be the dihedral angle between AOE and OAB. The faces AOB and AOE will intersect at $\varphi = 0$ and $\varphi = 2\pi$, so these are excluded from the configuration space.

The second degree of freedom, $\alpha$, is defined to be the dihedral angle from face AOE to DOE. It is symmetric, meaning it is also the dihedral angle from face AOB to face BOC. We chose this symmetry because we are examining the equilibrium pathways of the system. As with $\varphi$, it should be valid from 0 to $2\pi$ since it self intersects at these points.

And thus, for a purely kinematic origami system with no springs attached, and not accounting for the intersection of faces OBC and ODE, we can write that

\begin{equation}
    \varphi \in (0,2\pi) \quad \alpha \in (0,2\pi)
\end{equation}

%% file: supp/figs/tikz/valid-beta-gamma.tex
\begin{tikzpicture}
    \begin{axis}
    [
        axis lines = left,
        legend style={legend pos=south east},
        xlabel = $\beta$,
        ylabel = $\gamma$,
        ymin = 0,
        xmax = pi
    ]

    % shaded region

    \draw[draw =  none, fill=gray!20] (0,3*pi/4) -- (pi/2,5*pi/8) -- (pi/2,3*pi/4) -- (0,pi) -- (0,3*pi/4) ;
    
    % blue line
    \addplot
    [
        domain=0:pi, 
        range = 0:pi,
        samples=100, 
        color = blue
    ]
    { pi-x/2 };
    \addlegendentry{$\gamma = \pi - \frac{\beta}{2}$}

    % orange line
    \addplot
    [
        domain=0:pi, 
        samples=100, 
        color = orange
    ]
    { 3*pi/4 - x/4 };
    \addlegendentry{$\gamma = \frac{3\pi}{4} - \frac{\beta}{4}$}

    % gamma = pi
    \addplot
    [
        dashed,
        domain=0:pi, 
        samples=100, 
    ]
    { pi/2 };
    \addlegendentry{$\gamma = \pi/2$}

    % beta = pi
    \addplot
    [
        dash dot,
        samples=50,
        smooth,
        domain=0:6,magenta
    ]
    coordinates
    {
        (pi/2,0)
        (pi/2,pi)
    };
    \addlegendentry{$\beta = \pi/2$}

    % \draw[fill=gray!50] plot[smooth, samples=100, domain=1:e] (\x,ln \x) -| (0,0) -- cycle;

    \end{axis}
\end{tikzpicture}

%% file: supp/appendicies/kinematics/spheres.tex
\begin{figure}
    \centering
    \includegraphics{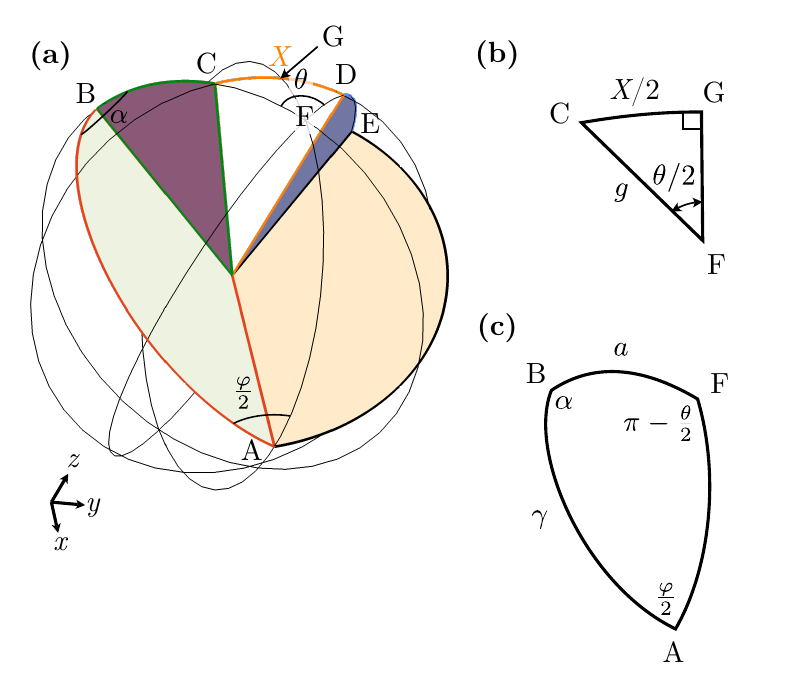}
    \caption{The System can Be Represented by a Spherical Trigonometry}
    \label{fig:spheres-top}
\end{figure}

If we assume that 
$\overset{\rightharpoonup}{OA}$, 
$\overset{\rightharpoonup}{OB}$,
$\overset{\rightharpoonup}{OC}$,
$\overset{\rightharpoonup}{OD}$,
and $\overset{\rightharpoonup}{OE}$ are all unit length, and that the point O stays fixed, then we can say that the points A, B, C, D, and E all lie on a sphere centered about point O (see Figure \ref{fig:spheres-top}a). This allows us to use spherical trigonometry to describe the kinematics.

With this viewpoint in mind, we define new points and arcs to create useful spherical triangles. First, if we extend the arcs BC and DE along the sphere, they meet at some point which we call $F$. Second, we define the midpoint of arc length $X$ to be point $G$. Finally, we define the arc that goes from point $C$ to point $F$ as $g$. This creates two triangles, CGF, and FAB shown in Figure \ref{fig:spheres-top}b and \ref{fig:spheres-top}c respectively.

%% file: supp/appendicies/kinematics/bounds.tex
There are two boundaries for this system not accounted for by our definitions of $\alpha$ and $\varphi$: (1) when faces OBC and ODE are parallel, and a rotational spring cannot be placed and (2) when the faces OBC and ODE intersect each other.

\subsubsection{Spring Boundary in Terms of $g$}
\begin{figure}
    \centering
    \includegraphics{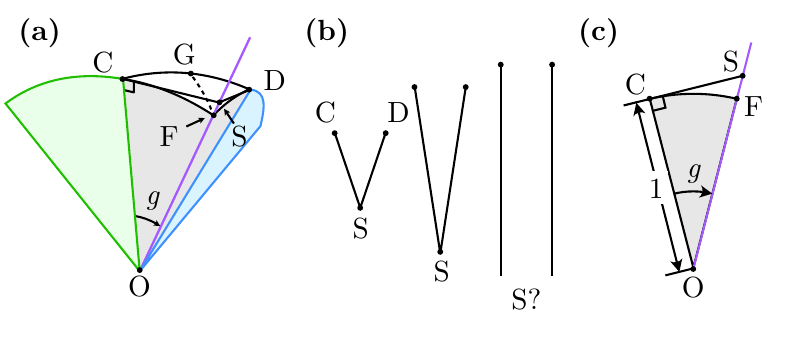}
    \caption{S is the intersection of CS and DS. When the meet at infinity, we define one boundary of the spring}
    \label{fig:bounds-trig}
\end{figure}

To understand the first boundary, imagine a rotational spring that is attached perpendicular to $\overset{\rightharpoonup}{OC}$ and $\overset{\rightharpoonup}{OD}$, and parallel to planes OBC and ODE respectively, which meet at some point S. This corresponds to a spring that lies along $\overset{\rightharpoonup}{CS}$ and $\overset{\rightharpoonup}{SD}$, as in Figure \ref{fig:bounds-trig}a.

There will be a point where the two lines, CS and DS meet at infinity, as shown in Figure \ref{fig:bounds-trig}b. Clearly, at this point, it is not possible to define a rotational spring across the faces. Notice that this is equivalent to the condition $|CS|\rightarrow\infty$

Using simple trigonometric analysis (Figure \ref{fig:bounds-trig}c), we can show that  

\begin{equation}
    |CS| = \tan g
\end{equation}

And when we take the limit, to find the bounds of the system, we get that

\begin{equation}
    \lim_{CS \rightarrow \infty} |CS| = \tan g_S
    \quad g_{S} = \frac{\pi}{2} \pm \pi n, n = 0,1,2 ...
    \label{eq:g-s-dfn}
\end{equation}

where $g_{S}$ is the kinematic limit of the rotational spring. 

\subsubsection{Intersect Boundary in Terms of $g$}

We now aim to find the second boundary -- the intersect boundary. Using the spherical law of sines on triangle GCF in Figure \ref{fig:spheres-top}b, we can write

\begin{equation}
    \frac{\sin g}{\sin(\pi/2)} = \frac{\sin(X/2)}{\sin(\theta/2)}
    \label{eq:sin-g}
\end{equation}

The faces OCE and ODE will intersect when $X=0$, so EQ\ref{eq:sin-g} becomes

\begin{equation}
    \quad \sin g_I = 0 \quad g_{I} = \pm\pi n,  n = 0,1,2 ...
\end{equation}

where $g_{I}$ is the kinematic limit of facet intersection.

\subsubsection{Interval of Validity}
\label{section:valid-interval}

\begin{figure}
    \centering
    \begin{subfigure}[b]{0.4\textwidth}
        \centering
        % first figure 
        \input{supp/figs/tikz/lower-g.tex}
        \caption{}
    \end{subfigure}
    \hfill
    \begin{subfigure}[b]{0.4\textwidth}
        \centering
        % second figure
        \input{figs/tikz/upper-g.tex}
        \caption{}
    \end{subfigure}
    \caption{The model is either valid on (a) $g\in(0,\frac{\pi}{2})$ or (b) $g\in(\frac{\pi}{2},\pi)$}
    \label{fig:region-question}
\end{figure}

\begin{figure}
    \centering
    \input{supp/figs/tikz/beta-g.tex}
    \caption{The test point $(\alpha,\varphi) = (\pi,\pi)$ is equivalent to $g = \beta$. Therefore, if $\beta \in(0,\frac{\pi}{2})$, then $g\in(0,\frac{\pi}{2})$}
    \label{fig:region-answer}
\end{figure}

We have now established the boundaries of the system in terms of $g$. In this section, we aim to determine what range of $g$ values correspond to the system we are investigating.

Taking equation \ref{eq:sin-g} in the general case, we see that 

\begin{equation}
    \sin g = \frac{\sin (X/2)}{\sin (\theta/2)}
    \label{eq:reduced-sin-g}
\end{equation}

and the Pythagorean identity tells us that 

\begin{equation}
    \cos g = \pm \sqrt{1 - \sin^2 g}
    \label{eq:two-solutions}
\end{equation}

indicating there are two possible kinematic spaces for a given $\sin g$. When we plug $g = \pi/2$ into EQ\ref{eq:two-solutions} , we find that 

\begin{equation}
    0 = \pm 0
\end{equation}

or that the two possible solutions overlap with each other at $g=\pi/2$, no matter what the parameters of the system are. 

We can go further to eliminate all negative values of $g$, since that would imply that the spring intersects with the facets. We can also eliminate all values of $g$ above $\pi$, since values above $\pi$ imply self intersection of the spring. 

Summarizing this, we can say that $g$ is always somewhere between 0 and $\pi$ and that it branches into one of two possible solution spaces at $g = \pi/2$. Another way of saying this is that the model is either valid on $g\in(0,\pi/2)$ or $g\in(\pi/2,\pi)$, as shown in Figure \ref{fig:region-question}.

To determine which branch we lie on, we can use a test point which we know is always in our configuration space, that is $(\alpha, \varphi) = (\pi,\pi)$, which is the flat state defined in Figure \ref{fig:geometry}. From this figure, it is simple to see that as $\theta \rightarrow \pi$ then $X\rightarrow g\rightarrow\beta$. Therefore EQ \ref{eq:reduced-sin-g} becomes 

\begin{equation}
    \sin (g_{flat}) = \sin \beta
\end{equation}

Where $g_{flat}$ is the value of $g$ when $(\alpha, \varphi) = (\pi,\pi)$. Since we know that $0 < g <\pi$, then $g_{flat} = \beta$. This implies that if $\beta < \pi/2$, we will be in the branch in Figure \ref{fig:region-question}a and if $\beta > \pi/2$ we will be in the branch in Figure \ref{fig:region-question}b, with the case of $\beta = \pi/2$ being ambiguous.

In this work, for simplicity, we will only consider $\beta \in (0,\pi/2)$, which implies that $g \in (0,\pi/2)$, as shown in Figure \ref{fig:region-answer}

\subsubsection{Spring Boundary in Terms of $\alpha$ and $\varphi$}

In the previous sections, we established that the system is valid from $g \in (0,\pi/2)$ where $g = 0$ is the facet intersection boundary and $g =\pi/2$ is the spring intersection boundary. In the following sections, we would like to write these boundaries in terms of $\varphi$ and $\alpha$. We will start with the spring boundary. 

We will begin by using the four part cotangent formula \cite{todhunter1863spherical} on triangle ABF (Figure \ref{fig:spheres-top}) to relate $\gamma$, $\alpha$, $\varphi$, and $a$

\begin{equation}
    \cos\gamma\cos\alpha = \cot a\sin\gamma - \cot\frac{\varphi}{2}\sin\alpha
\end{equation}

rearranging, we get

\begin{equation}
    \cot \frac{\varphi}{2} = \frac{\cot a \sin \gamma - \cos \gamma \cos \alpha}{\sin \alpha}
\end{equation}

or 

\begin{equation}
     \tan\left(\frac{\pi}{2}- \frac{\varphi}{2}\right) = \frac{\cot a \sin \gamma - \cos \gamma \cos \alpha}{\sin \alpha}
    \label{eq:tan-shift}
\end{equation}

or 

\begin{equation}
     \tan\frac{\varphi}{2} = \frac{\sin \alpha}{\cot a \sin \gamma - \cos \gamma \cos \alpha}
    \label{eq:tan-no-shift}
\end{equation}

From Figure \ref{fig:spheres-top} we can write $a$ in terms of $g$ as

\begin{equation}
    a = g + \eta
    \label{eq:a-base}
\end{equation}

In  general $\eta = \pi - \gamma - \beta/2$, as in EQ \ref{eq:eta}. For the case where we are at the spring boundary ($g = g_S$), then  $g\rightarrow\pi/2$ and . Substituting these into EQ \ref{eq:a-base}, we get

\begin{equation}
    a_S = \frac{3\pi}{2} - (\gamma +\beta/2)
\end{equation}

where $a_S$ is the value of $a$ at the spring boundary.
Taking the cotangent of both sides, to match the from in EQ \ref{eq:tan-shift}, we get

\begin{equation}
    \cot a_S = \cot \left(\frac{3\pi}{2} - (\gamma +\beta/2)\right) = \tan (\gamma + \beta/2)
\end{equation}

inserting this into EQ \ref{eq:tan-shift} and rearranging for $\varphi$, we get the following

\begin{equation}
    \varphi_{S} (\alpha) = 
    \pi - 2 \tan ^{-1}
    \Bigg( 
    \frac{\tan(\gamma + \beta/2) \sin \gamma - \cos\gamma \cos\alpha}{\sin \alpha}
    \Bigg)
\end{equation}

where $\varphi_S(\alpha)$ is the spring boundary in terms of  $\alpha$.

\subsubsection{Intersect Boundary in Terms of $\alpha$ and $\varphi$}

Now we aim to find an equation for the second boundary. If we instead plug $g = 0$, representing the condition of the intersect boudary, into EQ \ref{eq:a-base}, we get that 

\begin{equation}
    a_I = \pi - (\gamma+\beta/2)
\end{equation}

where $a_I$ is the value of $a$ at the intersect boundary, and so $\cot a$ becomes

\begin{equation}
    \cot a_I = \cot\left(\pi - (\gamma+\beta/2)\right) =  - \cot (\gamma + \beta/2)
\end{equation}

and so the boundary becomes 

\begin{equation}
    \quad
    \varphi_{I}(\alpha) = 
    2 \tan ^{-1}
    \Bigg( 
    \frac{\sin \alpha}{-\cot(\gamma + \beta/2) \sin \gamma - \cos\gamma \cos\alpha}
    \Bigg) + 2\pi n
    \quad 
    n = 0,1
    \label{eq:interset-dfn}
\end{equation}

where $\varphi_I(\alpha)$ is the intersect boundary in terms of  $\alpha$, and $n$ is adjusted to keep $\varphi$ between 0 and $2\pi$.

Notice that the condition $x=0$ is equivalent to the non-crease stretching solution, so this same equation represents the Non-Euclidean system, and the boundaries of this system.

%% file: supp/figs/tikz/lower-g.tex
\begin{tikzpicture}

    % define coordinates
    \coordinate (a) at (0,0);
    \coordinate (b) at (4,0);
    \coordinate (yMargin) at (0,0.5);
    \coordinate (xMargin) at (0.5,0);

    % labels
    \node at ($(a) - (yMargin)$) {$g_I = 0$};
    \node at ($(b) - (yMargin)$) {$g_S = \frac{\pi}{2}$};

    % number line
    \draw[latex-latex] ($(a) - (xMargin)$) -- ($(b) + (xMargin)$);

    % thick line
    \draw[very thick] (a) -- (b);
    
    % nodes
    \node[circle,draw=black, fill=white, inner sep=0pt,minimum size=5pt] at (a) {};
    \node[circle,draw=black, fill=white, inner sep=0pt,minimum size=5pt] at (b) {};

\end{tikzpicture}

%% file: supp/figs/tikz/upper-g.tex
\begin{tikzpicture}

    % define coordinates
    \coordinate (a) at (0,0);
    \coordinate (b) at (4,0);
    \coordinate (yMargin) at (0,0.5);
    \coordinate (xMargin) at (0.5,0);

    % labels
    \node at ($(a) - (yMargin)$) {$g_S = \frac{\pi}{2}$};
    \node at ($(b) - (yMargin)$) {$g_I = \pi$};

    % number line
    \draw[latex-latex] ($(a) - (xMargin)$) -- ($(b) + (xMargin)$);

    % thick line
    \draw[very thick] (a) -- (b);
    
    % nodes
    \node[circle,draw=black, fill=white, inner sep=0pt,minimum size=5pt] at (a) {};
    \node[circle,draw=black, fill=white, inner sep=0pt,minimum size=5pt] at (b) {};

\end{tikzpicture}

%% file: supp/figs/tikz/beta-g.tex
\begin{tikzpicture}

    % define coordinates
    \coordinate (a) at (0,0);
    \coordinate (b) at (4,0);
    \coordinate (beta) at (1.2,0);
    \coordinate (yMargin) at (0,0.5);
    \coordinate (xMargin) at (0.5,0);

    % labels
    \node at ($(a) - (yMargin)$) {$g_I = 0$};
    \node at ($(b) - (yMargin)$) {$g_S = \frac{\pi}{2}$};
    \node at ($(beta) - (yMargin)$) {$\beta$};
    \node at ($(beta) + (yMargin)$) {$\alpha = \varphi = \pi$};

    % number line
    \draw[latex-latex] ($(a) - (xMargin)$) -- ($(b) + (xMargin)$);

    % thick line
    \draw[very thick] (a) -- (b);
    
    % nodes
    \node[circle,draw=black, fill=white, inner sep=0pt,minimum size=5pt] at (a) {};
    \node[circle,draw=black, fill=white, inner sep=0pt,minimum size=5pt] at (b) {};
    \node[circle,draw=white, fill=black, inner sep=0pt,minimum size=5pt] at (beta) {};

\end{tikzpicture}

%% file: supp/appendicies/kinematics/closed-region.tex
\begin{figure}
    \centering
    \input{figs/tikz/bounded-phi-alpha.tex}
    \caption{$\varphi_I(\alpha)$ and $\varphi_S(\alpha)$ form a closed region, if points A,B,C and D are included in the boundary. They always approach the points shown in this graph and are always continuous between these points if the constraints in Figure \ref{fig:param-limits} are satisfied.}
    \label{fig:bounded-phi-alpha}
\end{figure}

In this section, we aim to show that the spring boundary and intersect boundary established in the previous section form a closed region, given the constraints established in Figure \ref{fig:param-limits}.

First, we will show that $\varphi_I(\alpha)$ and $\varphi_S(\alpha)$ always approach the endpoints (A,B,C, or D) that are shown in Figure \ref{fig:bounded-phi-alpha}. Then we will show that the $\varphi_I(\alpha)$ and $\varphi_S(\alpha)$ are always continuous between these points, and the only discontinuities occur at the endpoints. Therefore including $\varphi_I(\alpha)$ and $\varphi_S(\alpha)$ along with points A,B,C and D create a boundary that is always closed given the constraints established in Figure \ref{fig:param-limits}.

\subsubsection{Limits on Point A}

\paragraph{Intersect Boundary Limit}

From Figure \ref{fig:bounded-phi-alpha}, we see that the intersect boundary approaches point A from the positive direction on the branch of $\varphi_I$ where $n = 0$. We can write this as

\begin{equation}
    \lim_{\alpha \to 0^+} \varphi_I(\alpha) = 
    \lim_{\alpha \to 0^+}
    2 \tan ^{-1}
    \Bigg( 
    \frac{\sin \alpha}{-\cot(\gamma + \beta/2) \sin \gamma - \cos\gamma \cos\alpha}
    \Bigg)
\end{equation}

which reduces to 

\begin{equation}
    \lim_{x \to 0}
    2 \tan ^{-1} (x) = \boxed{0}
\end{equation}

\paragraph{Spring Boundary Limit}
\label{section:lower-limit-justify}

In this section, we will justify the condition that $\gamma>\frac{3\pi}{4} - \frac{\beta}{4}$. This starts from enforcing that the spring boundary also approaches point A in Figure \ref{fig:bounded-phi-alpha}. Mathematically, we can write this as 

\begin{equation}
    \lim_{\alpha \to 0^+} \varphi_{S} (\alpha) = 0
\end{equation}

Expanding the left side, we can write 

\begin{equation}
    \lim_{\alpha \to 0^+}
    \varphi_{S} (\alpha) = 
    \lim_{\alpha \to 0^+} \pi - 2 \tan ^{-1}
    \Bigg( 
    \frac{\tan(\gamma + \beta/2) \sin \gamma - \cos\gamma \cos\alpha}{\sin \alpha}
    \Bigg) = 0
\end{equation}

which can be rewritten as

\begin{equation}
    \pi - \lim_{x \to 0^+} 2 \tan ^{-1}
    \Bigg( 
    \frac{\tan(\gamma + \beta/2) \sin \gamma - \cos\gamma }{x}
    \Bigg) = 0
    \label{eq:a-spring-limit-base}
\end{equation}

to simplify the numerator, we can rewrite

\begin{equation}
    \tan(\gamma + \beta/2) \sin \gamma - \cos\gamma   
\end{equation}

as

\begin{equation}
     \frac{\sin(\gamma + \beta/2) \sin \gamma - \cos(\gamma + \beta/2)\cos\gamma}{\cos(\gamma + \beta/2)}
\end{equation}

and by using the cosine angle addition identities, we get can rewrite this as 

\begin{equation}
    \tan(\gamma + \beta/2) \sin \gamma - \cos\gamma  = -\frac{\cos(2\gamma + \beta/2)}{\cos(\beta/2 + \gamma)}
    \label{eq:ratio}
\end{equation}

if we substitute EQ \ref{eq:ratio} into EQ \ref{eq:a-spring-limit-base}, we get that 

\begin{equation}
    \pi - \lim_{x \to 0^+} 2 \tan ^{-1}
    \left( 
        -\frac{1}{x}
        \frac{\cos(2\gamma + \beta/2)}{\cos(\beta/2 + \gamma)}
    \right) = 0
\end{equation}

this is equivalent to writing that 

\begin{equation}
    \lim_{x \to 0^+} 
    \left( 
        -\frac{1}{x}
        \frac{\cos(2\gamma + \beta/2)}{\cos(\beta/2 + \gamma)}
    \right) \rightarrow +\infty
\end{equation}

or that 

\begin{equation}
    -\frac{\cos(2\gamma + \beta/2)}{\cos(\beta/2 + \gamma)} > 0
    \label{eq:ratio-constrained}
\end{equation}

we established in EQ \ref{eq:beta-gamma-constraints} that $ \pi/2 < \beta/2 + \gamma < \pi$, therefore $\cos(\beta/2 + \gamma) < 0$. Thus EQ \ref{eq:ratio-constrained} reduces to 

\begin{equation}
    \cos(2\gamma + \beta/2) > 0
\end{equation}

which can also be written as

\begin{equation}
    -\frac{\pi}{2} + 2\pi n < 2\gamma + \frac{\beta}{2} < \frac{\pi}{2} + 2\pi n\quad n = ... -1, 0, 1 ...
    \label{eq:n-bounds}
\end{equation}

from EQ \ref{eq:base-limits}, we know that the minimum value that $2\gamma + \beta/2$ could have is $\pi$, and the maximum is $\frac{5\pi}{2}$. This means we are in the $n=2$ branch of  EQ \ref{eq:n-bounds}. So it can be rewritten as

\begin{equation}
    \frac{3\pi}{2} < 2\gamma + \beta/2 < \frac{5\pi}{2}
\end{equation}

We can rewrite 

\begin{equation}
    \frac{3\pi}{2} < 2\gamma + \beta/2 
\end{equation}

as 

\begin{equation}
    \gamma > 3\pi/4 -\beta/4
\end{equation}

which is the constraint we used in EQ \ref{eq:lower-gamma}, and which ensures that 
\begin{equation}
    \lim_{\alpha \to 0^+} \varphi_{S} (\alpha) = \boxed{0}
\end{equation}

\subsubsection{Limits on Point B}

\paragraph{Intersect Boundary Limit}

From Figure \ref{fig:bounded-phi-alpha}, we see that the intersect boundary approaches point B from the negative direction on the branch of $\varphi_I$ where $n = 0$. We can write this as

\begin{equation}
    \lim_{\alpha \to \pi^-} \varphi_I(\alpha) = 
    \lim_{\alpha \to \pi^-}
    2 \tan ^{-1}
    \Bigg( 
    \frac{\sin \alpha}{-\cot(\gamma + \beta/2) \sin \gamma - \cos\gamma \cos\alpha}
    \Bigg)
\end{equation}

which reduces to 

\begin{equation}
    \lim_{x \to 0}
    2 \tan ^{-1} (x) = \boxed{0}
\end{equation}

\paragraph{Spring Boundary Limit}

From Figure \ref{fig:bounded-phi-alpha}, we see that the spring boundary approaches point B from the positive direction. We can write this as

\begin{equation}
    \lim_{\alpha \to \pi^+}
    \varphi_{S} (\alpha) = 
    \lim_{\alpha \to \pi^+} \pi - 2 \tan ^{-1}
    \Bigg( 
    \frac{\tan(\gamma + \beta/2) \sin \gamma - \cos\gamma \cos\alpha}{\sin \alpha}
    \Bigg)
\end{equation}

which can be rewritten as 

\begin{equation}
    \pi - \lim_{x \to 0^-} 2 \tan ^{-1}
    \Bigg( 
    \frac{\tan(\gamma + \beta/2) \sin \gamma + \cos\gamma}{x}
    \Bigg)
    \label{eq:b-intermediate-limit}
\end{equation}

if we apply the constraints from EQ \ref{eq:base-limits} and EQ \ref{eq:beta-gamma-constraints} to the numerator, that is that $\gamma \in (\pi/2,\pi)$ and $\pi/2<\gamma + \beta/2 < \pi$, we see that

\begin{equation}
    \tan(\gamma+\beta/2) < 0 \quad \sin\gamma > 0 \quad \cos\gamma < 0
\end{equation}

and therefore 

\begin{equation}
    \tan(\gamma + \beta/2) \sin \gamma + \cos\gamma < 0
    \label{eq:sum-bounds}
\end{equation}

which, if $\gamma \in (\pi/2, \pi)$, can be written as 

\begin{equation}
    \pi - \lim_{x \to \infty}
    2 \tan ^{-1} (x) = \boxed{0}
\end{equation}

\subsubsection{Limits on Point C}

\paragraph{Intersect Boundary Limit}

From Figure \ref{fig:bounded-phi-alpha}, we see that the intersect boundary approaches point C from the positive direction on the branch of $\varphi_I$ where $n = 1$. We can write this as

\begin{equation}
    \lim_{\alpha \to \pi^+} \varphi_I(\alpha) = 
    \lim_{\alpha \to \pi^+}
    2 \tan ^{-1}
    \Bigg( 
    \frac{\sin \alpha}{-\cot(\gamma + \beta/2) \sin \gamma - \cos\gamma \cos\alpha}
    \Bigg) + 2\pi
\end{equation}

which reduces to 

\begin{equation}
    2\pi + \lim_{x \to 0}
    2 \tan ^{-1} (x) = \boxed{2\pi}
\end{equation}

\paragraph{Spring Boundary Limit}

From Figure \ref{fig:bounded-phi-alpha}, we see that the spring boundary approaches point C from the negative direction. We can write this as

\begin{equation}
    \lim_{\alpha \to \pi^-}
    \varphi_{S} (\alpha) = 
    \lim_{\alpha \to \pi^-} \pi - 2 \tan ^{-1}
    \Bigg( 
    \frac{\tan(\gamma + \beta/2) \sin \gamma - \cos\gamma \cos\alpha}{\sin \alpha}
    \Bigg)
\end{equation}

which can be rewritten as 

\begin{equation}
    \pi - \lim_{x \to 0^+} 2 \tan ^{-1}
    \Bigg( 
    \frac{\tan(\gamma + \beta/2) \sin \gamma + \cos\gamma}{x}
    \Bigg)
    \label{eq:intermediate-limit}
\end{equation}

from EQ \ref{eq:sum-bounds}, we know that

\begin{equation}
    \tan(\gamma + \beta/2) \sin \gamma + \cos\gamma < 0
\end{equation}

and thus EQ \ref{eq:intermediate-limit} becomes

\begin{equation}
    \pi - \lim_{y \to -\infty}
    2 \tan ^{-1} (y) = \boxed{2\pi}
\end{equation}

\subsubsection{Limits on Point D}

\paragraph{Intersect Boundary Limit}

From Figure \ref{fig:bounded-phi-alpha}, we see that the intersect boundary approaches point D from the negative direction on the branch of $\varphi_I$ where $n = 1$. We can write this as

\begin{equation}
    \lim_{\alpha \to 2\pi^-} \varphi_I(\alpha) = 
    \lim_{\alpha \to 2\pi^-}
    2 \tan ^{-1}
    \Bigg( 
    \frac{\sin \alpha}{-\cot(\gamma + \beta/2) \sin \gamma - \cos\gamma \cos\alpha}
    \Bigg) + 2\pi
\end{equation}

which reduces to 

\begin{equation}
    2\pi + \lim_{x \to 0}
    2 \tan ^{-1} (x) = \boxed{2\pi}
\end{equation}

\paragraph{Spring Boundary Limit}

From Figure \ref{fig:bounded-phi-alpha}, we see that the spring boundary approaches point D from the negative direction. We can write this as

\begin{equation}
    \lim_{\alpha \to 2\pi^-}
    \varphi_{S} (\alpha) = 
    \lim_{\alpha \to 2\pi^-} \pi - 2 \tan ^{-1}
    \Bigg( 
    \frac{\tan(\gamma + \beta/2) \sin \gamma - \cos\gamma \cos\alpha}{\sin \alpha}
    \Bigg)
\end{equation}

which reduces to 

\begin{equation}
    \pi - \lim_{x \to 0^-} 2 \tan ^{-1}
    \left( 
    -
    \frac{1}{x}
    \frac{\cos(2\gamma + \beta/2)}{\cos(\beta/2 + \gamma)}
    \right)
    \label{eq:d-limit}
\end{equation}

from EQ \ref{eq:ratio-constrained}, we know that 

\begin{equation}
    -\frac{\cos(2\gamma + \beta/2)}{\cos(\beta/2 + \gamma)} > 0
\end{equation}

and so EQ \ref{eq:d-limit} reduces to 

\begin{equation}
    \pi - \lim_{y \to -\infty}
    2 \tan ^{-1} (y) = \boxed{2\pi}
\end{equation}

\subsubsection{Continuity of A$\rightarrow$C and B$\rightarrow$D (Spring Boundary)}

In this section, we aim to show that the only discontinuities of the Spring boundary are at points A,B,C and D, and it is continuous everywhere else.

Below, is the equation for the spring boundary

\begin{equation}
    \varphi_{S} (\alpha) = 
    \pi - 2 \tan ^{-1}
    \Bigg( 
    \frac{\tan(\gamma + \beta/2) \sin \gamma - \cos\gamma \cos\alpha}{\sin \alpha}
    \Bigg)
\end{equation}

It's only discontinuities are when 

\begin{equation}
    \sin\alpha = 0 
    \label{eq:sin-alpha}
\end{equation}

which is true when $\alpha = 0$ or $\alpha = \pi$ or $\alpha = 2\pi$, which correspond to points A, B and C, and finally D, respectively.

\subsubsection{Continuity of A$\rightarrow$B and C$\rightarrow$D (Intersect Boundary)}

In this section, we aim to show that the intersect boundary is never discontinous.

\begin{equation}
    \quad
    \varphi_{I}(\alpha) = 
    2 \tan ^{-1}
    \Bigg( 
    \frac{\sin \alpha}{-\cot(\gamma + \beta/2) \sin \gamma - \cos\gamma \cos\alpha}
    \Bigg) + 2\pi n
    \quad 
    n = 0,1
\end{equation}

It's only discontinuities are when 

\begin{equation}
    -\cot(\gamma + \beta/2) \sin \gamma - \cos\gamma \cos\alpha = 0
\end{equation}

which is equivalent to writing that 

\begin{equation}
    \frac{-\cot(\gamma + \beta/2)}{\cot\gamma} = \cos\alpha
\end{equation}

since $\cos\alpha \in [-1,1]$, if we show that $-\cot(\gamma+\beta/2)/\cot\gamma < -1$, then we can show that $\varphi_I$ has no discontinuities. 

We start with the bounds established in EQ \ref{eq:base-limits} and EQ \ref{eq:beta-gamma-constraints}, that

\begin{equation}
    \pi/2 < \gamma + \beta/2 < \pi \quad \textrm{and} \quad \pi/2 < \gamma < \pi
\end{equation}

Taking the cotangent of both of these, we see that 

\begin{equation}
    \cot(\gamma + \beta/2) < 0 \quad \textrm{and} \quad \cot\gamma < 0
\end{equation}

since we know they are both negative, and that $\beta > 0$, we can say that 

\begin{equation}
    \cot(\gamma + \beta/2) < \cot\gamma
\end{equation}

which, since $\cot\gamma <0$, becomes 

\begin{equation}
    \frac{\cot(\gamma + \beta/2)}{\cot\gamma} > 1
\end{equation}

multiplying by a negative, we get that 

\begin{equation}
    \frac{-\cot(\gamma + \beta/2)}{\cot\gamma} < -1
\end{equation}

and so 

\begin{equation}
    \frac{-\cot(\gamma + \beta/2)}{\cot\gamma} \neq \cos\alpha
\end{equation}

thus

\begin{equation}
    -\cot(\gamma + \beta/2) \sin \gamma - \cos\gamma \cos\alpha \neq 0
\end{equation}

therefore $\varphi_I(\alpha)$ is never discontinuous.

%% file: supp/figs/tikz/bounded-phi-alpha.tex
\begin{tikzpicture}

    \tikzmath{
        real \b, \g;
        \b = 5;             % degree
        \g  = deg(3*pi/4) ;  % degree
    }
    
    \begin{axis}[
        legend style={legend pos=north west},
        xlabel = {$\alpha$},
        ylabel = {$\varphi$},
    ]
    
    % lower
    
    \addplot
    [
        densely dashed,
        domain=0:pi, 
        samples=100, 
        color=blue,
    ]
    {
        rad(2*atan(-sin(deg(x))/(cot(\g +\b/2)*sin(\g)+ cos(\g)*cos(deg(x)))))
    };

    % upper
    
    \addplot
    [
        loosely dashed,
        domain=pi:2*pi, 
        samples=100, 
        color=blue,
    ]
    {
        2*pi + rad(2*atan(-sin(deg(x))/(cot(\g +\b/2)*sin(\g)+ cos(\g)*cos(deg(x)))))
    };

    % left
    \addplot
    [
        densely dotted,
        domain=0:pi, 
        samples=100, 
        color=blue,
    ]
    {
        pi - rad(2*atan((tan(\g +\b/2)*sin(\g)- cos(\g)*cos(deg(x)))/sin(deg(x))))
    };

    % right

    \addplot
    [
        densely dotted,
        domain=pi+0.01:2*pi, 
        samples=100, 
        color=blue,
    ]
    {
        pi - rad(2*atan((tan(\g +\b/2)*sin(\g)- cos(\g)*cos(deg(x)))/sin(deg(x))))
    };
    
    \addlegendentry{$\varphi_I, n = 0$}
    \addlegendentry{$\varphi_I, n = 1$}
    \addlegendentry{$\varphi_S$}

    \draw (pi,pi) node[] {valid regime (closed)};
    
    % \fill[red] (0,0) circle (3pt) node[below left = 2pt] {A};
    % \fill[red] (pi,0) circle (3pt) node[below right = 2pt] {B};

    \addplot[draw=blue!0,mark=*,mark options={color=blue!70}] coordinates {(0,0) (pi,0) (pi,2*pi) (2*pi,2*pi)};

    % \draw (0,0) node[] {A}

    \draw (0,0) node[below right] {A: (0,0)};
    \draw (pi,0) node[below right] {B: ($\pi$,0)};
    \draw (pi,2*pi) node[below ] {C: ($\pi$,$2\pi$)};
    \draw (2*pi,2*pi) node[below left] {D: ($2\pi$,$2\pi$)};

    \end{axis}

    % \draw (0,0)  
    
    % node[size=0.2cm,draw=red,fill= red,circle,dashed,label ={[red] below:$A:{0,0}$}] {};
    % \draw (0,0)  node[minimum size=1cm,draw=red,circle,label = below:$A:{0,0}$] {};

\end{tikzpicture}

%% file: supp/appendicies/kinematics/x-theta.tex
In this section, we will show that each $x-\theta$ pair corresponds to two $\alpha-\varphi$ pairs. First, we will define $x$ and $\theta$ more precisely. As shown in Figure \ref{fig:DOF} there are two energetic elements in the basic system: the rotational spring and the extensional spring. Simply put, $x$ is the euclidean distance between points C and D, and tracks the extensional spring (stiffness $k_E$), and $\theta$ is the angle between facets OBC and ODE and tracks the rotational spring (stiffness $k_T$). Mathematically, this can be written as 

\begin{equation}
    U = \frac{1}{2} \bigg(k_E(x-x_0)^2 + k_T(\theta-\theta_0)^2\bigg)
    \label{eq:energy}
\end{equation}

where $U$ is the energy of the system, and $x_0$ and $\theta_0$ are the equilibrium values for the extensional and rotational springs respectively.

\subsubsection{Two $\varphi$ values correspond to a given $x-\theta$ pair}

This section will focus on $\varphi$, and we will show that a unique $x-\theta$ pair will correspond to two values of $\varphi$, one which is $\varphi>\pi$, one which is $\varphi<\pi$. We will also show that at $\varphi=\pi$, the solutions overlap and become one.

\begin{figure}
    \centering
    \input{supp/figs/tikz/triangle.tex}
    \caption{Because OC and OD are unit length, the arc $\overset{\huge\frown}{DC}$ is the same as angle COD}
    \label{fig:x-X}
\end{figure}

To begin this analysis, we will note that since $x$ is the euclidean distance between $C$ and $D$, $X$ is the arc length from $C$ to $D$, and $\overset{\rightharpoonup}{OC}$ and $\overset{\rightharpoonup}{OD}$ are unit length, as in Figure \ref{fig:x-X}, then $x$ and $X$ can be related as follows

\begin{equation}
    \frac{x}{2} = \sin(X/2)
    \label{eq:little-x-big-x}
\end{equation}

substituting this into EQ \ref{eq:reduced-sin-g}, we get

\begin{equation}
    \sin g = \frac{x}{2}\csc(\theta/2)
    \label{eq:sin-g-little-x}
\end{equation}

from Appendix \ref{section:valid-interval}, we know that g will always be between 0 and $\pi/2$ if $\beta$ is between 0 and $\pi/2$. Thus, we know that $\cos g$ will always be positive. Therefore, we know that $g$ will always be in the range of $\sin^{-1}$ function. Thus we can write

\begin{equation}
    \quad g(x,\theta) = \sin^{-1}\bigg[\frac{x}{2} \csc\bigg(\frac{\theta}{2}\bigg)\bigg] \quad\forall \quad\beta\in(0,\pi/2)
    \label{eq:g-of-x-and-theta}
\end{equation}

Now, using the spherical law of sines on triangle FAB in Figure \ref{fig:spheres-top}, we say that

\begin{equation}
\label{eq:sine-law-phi-1}
\frac{\sin(\varphi/2)}{\sin(\eta + g)} = \frac{\sin(\pi-\theta/2)}{\sin(\gamma)}
\end{equation}

which can be rewritten as 

\begin{equation}
\varphi_1 (x,\theta) = 2\sin^{-1}\left[\frac{\sin(\theta/2)}{\sin(\gamma)}\sin(\eta+g(x,\theta))\right]
\end{equation}

and plugging in EQ \ref{eq:g-of-x-and-theta} and EQ \ref{eq:eta} and reducing, we get

\begin{equation}
\varphi_1
= 
2\sin^{-1}
\left[
    \frac{\sin(\theta/2)}{\sin\gamma}
    \sin
    \left(
        \gamma +
        \frac{\beta}{2} 
        - \sin^{-1}
        \left[
            \frac{x}{2\sin(\theta/2)} 
        \right]
    \right)
\right]
\quad
\varphi_1 \in (0,\pi)
\label{eq:phi-1}
\end{equation}

\begin{figure}
    \centering
    \includegraphics{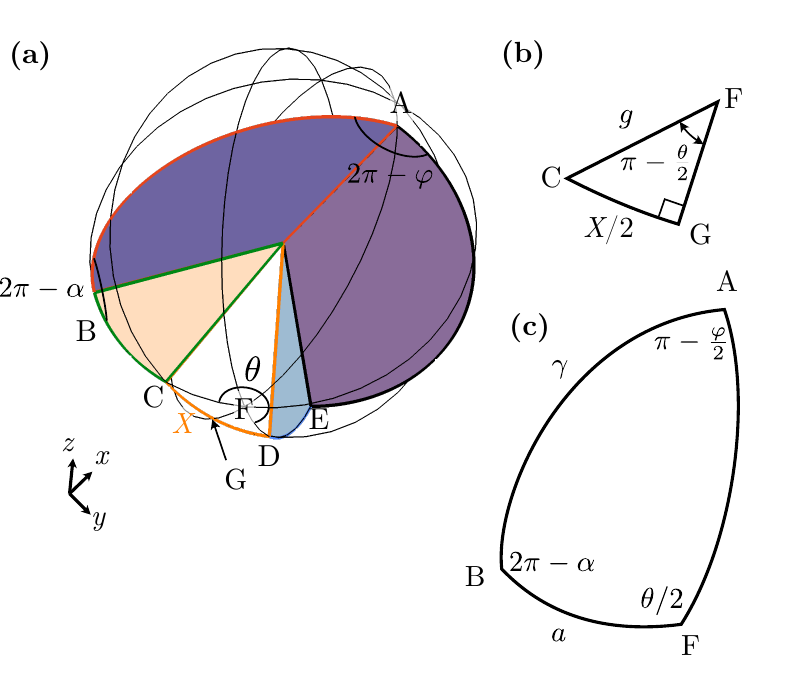}
    \caption{A different set of spherical triangles exists for $\varphi>\pi$}
    \label{fig:spheres-bot}
\end{figure}

For this equation we exclude values of $\varphi<0$ because, as discussed in Appendix \ref{section:parms-DOF} that would imply self-intersection of the facets. We also exclude values of $\varphi > \pi$ in this equation because $\sin^{-1}$ does not produce values above $\pi/2$ and so $2\sin^{-1}$ will not produce values above $\pi$. Therefore, we need another equation to describe the behavior of the system for $\varphi >\pi$. We turn to Figure \ref{fig:spheres-bot}, which shows how the spherical triangles change for $\varphi > \pi$

Using the spherical law of sines on triangle FAB in Figure \ref{fig:spheres-bot}, we get

\begin{equation}
\label{eq:sine-law-phi-2}
\frac{\sin(\pi -\varphi/2)}{\sin(\eta + g)} = \frac{\sin(\theta/2)}{\sin(\gamma)}
\end{equation}

which when rearranged can be developed into the following expression

\begin{equation}
\varphi_2 = 2\pi - 2\sin^{-1}\left[\frac{\sin(\theta/2)}{\sin(\gamma)}\sin(\eta+g(x,\theta))\right]
\label{eq:phi-2}
\end{equation}

comparing this to EQ \ref{eq:phi-1}, we see that

\begin{equation}
\varphi_2 = 2\pi - \varphi_1
\quad
\varphi_2 \in (\pi,2\pi)
\end{equation}

In conclusion, we see that each $x-\theta$ pair produces one value of $g$, but each value of $g$ yields two different solutions for $\varphi$ which are symmetric about $\varphi =\pi$. Notice this implies that at $\varphi=\pi$, the two solutions collapse to one.

\subsubsection{Two $\alpha$ values correspond to a given $x-\theta$ pair}

In the previous section, we showed that two different inverse kinematic solutions exist, and that they change at $\varphi = \pi$. In this section, we extend this analysis to the second degree of freedom to obtain two expressions for $\alpha(x,\theta)$.

We begin, as with the $\varphi$ analysis, in the regime where $\varphi < \pi$, shown in Figure \ref{fig:spheres-top}. Using Napier's analogies \cite{todhunter1863spherical} on triangle FAB, we get the following 

\begin{equation}
    \label{eq:napier_1}
    \cot
    \frac{\alpha_1}{2} =  
    \frac{
        \cos\left(\frac{1}{2}[\gamma + \eta]\right)
    }{
        \cos\left(\frac{1}{2}[\gamma - \eta]\right)
    } \tan\left(\frac{\pi}{2} + \left[\frac{\varphi_1 (x,\theta)}{4} - \frac{\theta}{4}\right]\right)
\end{equation}

where $\alpha_1$ is values of $\alpha$ where $\varphi >\pi$
which can be rewritten as

\begin{equation}
    \alpha_1 = 2 \tan^{-1}\left[
        -\frac{
        \cos\left(\frac{1}{2}[\gamma + \eta]\right)
    }{
        \cos\left(\frac{1}{2}[\gamma - \eta]\right)
    } \tan\left(\frac{\varphi_1(x,\theta)}{4} - \frac{\theta}{4}\right)
    \right] + 2\pi n \quad \varphi_1 \in (0,\pi)
    \label{eq:alpha-1}
\end{equation}

where n is an integer $(0,1,2...)$, which we adjust to make sure that $\alpha$ stays between 0 and 2$\pi$. 

For the solution of $\alpha$ for $\varphi > \pi$, we turn now to Figure \ref{fig:spheres-bot}, and again use Napier's analogies on triangle FAB and reduce, which yields

\begin{equation}
    \alpha_2 = 2 \tan^{-1}\left[
        -\frac{
            \cos\left(\frac{1}{2}[\gamma - \eta]\right)
        }{
            \cos\left(\frac{1}{2}[\gamma + \eta]\right)
        } \cot\left(\frac{\varphi_2(x,\theta)}{4} + \frac{\theta}{4}\right)
    \right] + 2\pi n \quad \varphi_2 \in (\pi,2\pi)
    \label{eq:alpha-2}
\end{equation}

where n is again an integer $(0,1,2...)$, which we adjust to make sure that $\alpha$ stays between 0 and 2$\pi$. 

\subsubsection{Converting $x-\theta$ pairs to $\varphi-\alpha$ pairs}

Let us say that we would like to convert $x = 0.25, \theta = 1.39211,\gamma = 3\pi/4, \beta = 10^{\circ}$ to the two $\alpha-\varphi$ pairs, as in the main text of the paper. First, we plug these values into EQ\ref{eq:phi-1}, and get that 

\begin{equation}
    \boxed{\varphi_1 = \pi/2}
\end{equation}

Inserting $\varphi_1$ into EQ\ref{eq:phi-2}, we get that

\begin{equation}
    \boxed{\varphi_2 = 3\pi/2}
\end{equation}

We then insert $\varphi_1$ into EQ\ref{eq:alpha-1}, recalling that $\eta = \pi - \beta/2 - \gamma$. Also, we set n to 0 so that $0<\alpha_1<2\pi$, and get

\begin{equation}
    \boxed{\alpha_1 = 1.09743}
\end{equation}

Finally, we insert $\varphi_2$ into EQ\ref{eq:alpha-2}, . Again, we set n to 0 so that $0<\alpha_1<2\pi$, and get

\begin{equation}
    \boxed{\alpha_2 = 3.00807}
\end{equation}

%% file: supp/figs/tikz/triangle.tex
\begin{tikzpicture}

    % define coordinates
    \coordinate (o) at (0,0);
    \coordinate (c) at (6,-2.5);
    \coordinate (d) at (6,2.5);

    \draw 
    (o) node[left]{$O$} 
    -- 
    (c) node[below]{$C$} 
    -- node[left]{$x$}
    (d) node[above]{$D$} 
    -- 
    cycle;
    \pic [draw, -,angle radius = 6.5cm, "$X$",right,angle eccentricity=1] {angle = c--o--d} ;
     \pic [draw, ->,angle radius = 1.55cm, fill = teal!30,"$X$", right,angle eccentricity=1] {angle = c--o--d} ;

    % [label=below:$O$] (o) -- (c) -- [label=above:$D$] (d) -- cycle;

% note: need to add the arc length from D to C of the angle X
\end{tikzpicture}

% [my angle/.style = {draw, fill=teal!30,
%                    angle radius=7mm, 
%                    angle eccentricity=1.1, 
%                    right, inner sep=1pt,
%                    font=\footnotesize} 
%                    ]

% \draw   (0,0) coordinate[label=below:$O$] (o) -- 
%         (4,-2) coordinate[label=below:$C$] (c) -- node[right] {$x$}
%         (4,2) coordinate[label=above:$D$] (d) -- cycle;
% \pic[my angle, "$X$"] {angle = c--o--d};

%% file: supp/appendicies/kinematics/alpha-phi.tex
\subsubsection{$\theta$ as a function of $\varphi$ and $\alpha$}
We start by finding $\theta$ as a function of $\varphi$ and $\alpha$. Using the spherical law of cosines on the triangle FAB in Figure \ref{fig:spheres-top}, we get

\begin{equation}
    \cos(\pi-\theta/2)= -\cos\alpha \cos\frac{\varphi}{2} + \sin\alpha\sin\frac{\varphi}{2}\cos\gamma
    \label{eq:almost_theta}
\end{equation}

which reduces to

\begin{equation}
\boxed{
    \theta(\alpha,\varphi) = 2\cos^{-1}\left(\cos\alpha \cos\frac{\varphi}{2} - \sin\alpha\sin\frac{\varphi}{2}\cos\gamma\right)}
    \label{eq:theta-of-alpha-and-phi}
\end{equation}

\subsubsection{$x$ as a function of $\varphi$ and $\alpha$}

Now, we find $x$ as a function of $\varphi$ and $\alpha$. From Figure \ref{fig:spheres-top}, we can see that

\begin{equation}
    g = a-\eta
\end{equation}

Applying the spherical law of sines to triangle GCF in Figure \ref{fig:spheres-bot}b and Figure \ref{fig:spheres-top}b yields the following:

\begin{equation}
    \frac{\sin(X/2)}{\sin(\theta/2)} = \sin(a-\eta)
\end{equation}

using the sine angle difference identity, we get
\begin{equation}
    \sin(X/2) = \sin(\theta/2)(\sin a \cos \eta - \cos a \sin \eta)
\end{equation}

Factoring out a $\sin a$, we get 

\begin{equation}
    \sin(X/2) = \sin(\theta/2)\sin a (\cos \eta - \cot a \sin \eta)
\end{equation}

Inserting EQ \ref{eq:little-x-big-x}, we find that

\begin{equation}
    x = 2\sin(\theta/2)\sin a (\cos \eta - \cot a \sin \eta)
    \label{eq:x_sine_law_almost_done}
\end{equation}

To get rid of the $\sin a$ term, we use the following identity

\begin{equation}
    \sin a = \pm \frac{1}{\sqrt{1+\cot^2 a}}
    \label{eq:base-sin-a}
\end{equation}

From Appendix \ref{section:parms-DOF} and Appendix \ref{section:valid-interval}, we know that

\begin{equation}
    \eta \in (0,\pi/2)\quad\textrm{and}\quad g \in (0,\pi/2)
\end{equation}

and since 

\begin{equation}
    a = g + \eta
\end{equation}

that implies that

\begin{equation}
    \sin a > 0
\end{equation}

and so EQ \ref{eq:base-sin-a} becomes

\begin{equation}
    \sin a = \frac{1}{\sqrt{1+\cot^2 a}}
    \label{eq:final-sin-a}
\end{equation}

Applying EQ \ref{eq:final-sin-a} to EQ \ref{eq:x_sine_law_almost_done}, we obtain the following relation:

\begin{equation}
\boxed{
    x = \frac{\sin(\theta/2)(\cos \eta - \cot a \sin \eta)}{\sqrt{1+\cot^2 a}}
    \label{eq:x-of-alpha-and-phi}}
\end{equation}

which gives us $x$ in terms of $\theta$ (EQ \ref{eq:theta-of-alpha-and-phi}) and $\cot a$. We now find the value of $\cot a$ as a function of $\alpha$ and $\varphi$. Using the cotangent four part formula on triangle FAB, we obtain the following

\begin{equation}
    \cos \gamma \cos \alpha = \cot a \sin \gamma - \cot\frac{\varphi}{2} \sin \alpha
\end{equation}

which can be rearranged to obtain

\begin{equation}
    \cot a (\alpha,\varphi) = \frac{\cos \gamma \cos \alpha + \cot \frac{\varphi}{2} \sin \alpha}{\sin \gamma}
    \label{eq:cot-a}
\end{equation}

Therefore combining EQ \ref{eq:x-of-alpha-and-phi} with  EQ \ref{eq:cot-a} and EQ \ref{eq:theta-of-alpha-and-phi} yields $x$ as a function of $\alpha$ and $\varphi$

%% file: supp/appendicies/kinematics/x-theta-phi.tex
%% note: there are some interesting equations here which we have commented out for the sake of brevity 
\subsubsection{Level Curve Function}

In this section, we write the kinematics of the system as a function of $x$, $\theta$, and $\varphi$ in the form $f(x,\theta,\varphi) = 0$. We must include $\varphi$ to fully define the kinematics, because as discussed in Appendix \ref{section:x-theta-symmetry}, $x$ and $\theta$ are symmetric about $\varphi = \pi$.

Using the spherical law of sines on Figure \ref{fig:spheres-top}, we can say that

\begin{equation}
    \frac{\sin(\pi-\theta/2)}{\sin\gamma}= \frac{\sin(\varphi/2)}{\sin(\eta+g)}
    \label{eq:base-x-theta-phi}
\end{equation}

which can be rewritten as

\begin{equation}
    \frac{\sin(\theta/2)}{\sin\gamma} - \frac{\sin(\varphi/2)}{\sin(\eta+g)} = 0
\end{equation}

substituting in EQ \ref{eq:g-of-x-and-theta}, we get that 

\begin{equation}
    f(x,\theta,\varphi) = \frac{\sin(\theta/2)}{\sin\gamma} - \frac{\sin(\varphi/2)}{\sin\left(\eta + \sin^{-1}\left[\frac{x}{2\sin(\theta/2)}\right]\right)} = 0
    \label{eq:f-x-theta-phi}
\end{equation}

\subsection{$\theta$ as a function of x and $\varphi$}

Rearranging EQ \ref{eq:base-x-theta-phi}, we get

\begin{equation}
    \sin(\theta/2)\sin(\eta + g) = \sin(\varphi/2)\sin\gamma
\end{equation}

Using the angle sum identity on $\sin(\eta + g)$, we get

\begin{equation}
    \sin(\theta/2)(\sin\eta\cos g + \cos\eta\sin g) = \sin(\varphi/2)\sin\gamma
    \label{eq:theta-design-sum}
\end{equation}

Assuming that $g$ is never zero, which we showed is true in Appendix \ref{section:valid-interval},  EQ\ref{eq:sin-g-little-x} can be rearranged as follows

\begin{equation}
    \sin(\theta/2) = \frac{x}{2\sin g}
\end{equation}

Inserting this into EQ\ref{eq:theta-design-sum}, we can conclude

\begin{equation}
    \frac{x}{2}\left( \sin\eta\cot g + \cos\eta \right) = \sin(\varphi/2)\sin\gamma
\end{equation}

Assuming $x\neq0$, which is the same as assuming that $g\neq0$, we can reduce our equation further to

\begin{equation}
    \cot g = \frac{\frac{2}{x}\sin(\varphi/2)\sin\gamma - \cos\eta}{\sin\eta}
\end{equation}

Again, conditional on $g\neq0$, this can be rewritten as

\begin{equation}
    \tan g = \frac{\sin\eta}{\frac{2}{x}\sin(\varphi/2)\sin\gamma - \cos\eta}
\end{equation}

or 

\begin{equation}
  g = \tan^{-1}(A(\varphi,x));\quad A(\varphi,x) = \frac{\sin\eta}{\frac{2}{x}\sin(\varphi/2)\sin\gamma - \cos\eta}
  \label{eq:A-definition}
\end{equation}

Assume $0<\theta<2\pi$, we can again rearrange EQ\ref{eq:sin-g}, to yield

\begin{equation}
    \sin(\theta/2) = \frac{x}{2\sin g}
\end{equation}

and solving for $\theta$

\begin{equation}
    \theta = 2 \sin^{-1}\left(\frac{x}{2}\csc(g)\right)\quad\textrm{or}\quad \theta = 2\pi - \sin^{-1}\left(\frac{x}{2}\csc(g)\right)
    \label{eq:branching-theta}
\end{equation}

Taking the cosecant of EQ\ref{eq:A-definition}, we get that

\begin{equation}
    \csc g = \csc\left(\tan^{-1}(A(\varphi,x))\right)
\end{equation}

which reduces to 

\begin{equation}
    \csc g = \frac{\sqrt{A^2(\varphi,x) +1}}{A(\varphi,x)}
\end{equation}

and so EQ\ref{eq:branching-theta} becomes 

\begin{equation}
\boxed{
    \theta = 2 \sin^{-1}\left(\frac{x}{2}\frac{\sqrt{A^2(\varphi,x) +1}}{A(\varphi,x)}\right)\quad\textrm{or}\quad \theta = 2\pi - \sin^{-1}\left(\frac{x}{2}\frac{\sqrt{A^2(\varphi,x) +1}}{A(\varphi,x)}\right)}
\end{equation}

where A is defined in EQ\ref{eq:branching-theta}. Further, inserting some equilibrium value $\varphi_0$ into EQ\ref{eq:branching-theta}, we get 

\begin{equation}
    A(\varphi_{0,1},x) = \frac{\sin\eta}{\frac{2}{x}\sin(\varphi_{0,1}/2)\sin\gamma - \cos\eta}
\end{equation}

if we insert some $\varphi_{0,2}$ into this equation, where

\begin{equation}
    \pi - \varphi_{0,2}/2 = \varphi_{0,1}/2
    \label{eq:phi_0_2_define}
\end{equation}

noting that $\sin(\pi - \varphi_{0,2}/2) = \sin(\varphi_{0,1}/2) $, we can show that 

\begin{equation}
    A(\varphi_{0,1},x) = A(\varphi_{0,2},x)
\end{equation}

In conclusion, we have shown that for each x-value, there are two $\varphi$ values that will create the same $A$-value and hence the same $\theta$. We have also shown that they are symmetric about $\varphi = \pi$. Finally, we have shown that there are two possible $\theta$ values that will result in a given $\varphi-x$-pair.

%% file: supp/appendicies/energetics/energetics-intro.tex
Here, we reproduce the potential energy of the system

\begin{equation}
    U = \frac{1}{2} \bigg(k_E(x-x_0)^2 + k_T(\theta-\theta_0)^2\bigg)
    \label{eq:energy-reproduced}
\end{equation}

Where $x$ is the euclidean distance between points C and D, and tracks the extensional spring (stiffness $k_E$), and $\theta$ is the angle between facets OBC and ODE and tracks the rotational spring (stiffness $k_T$). In addition, $x_0$ and $\theta_0$ are the equilibrium values for the extensional and rotational springs respectively. The definitions of $x$ and $\theta$ are discussed in more detail in Appendix \ref{section:x-theta-symmetry}.

From \cite{li2020theory}, we know that all local minima of a system occur when the normal of the kinematic space ($\nabla f$, EQ \ref{eq:f-x-theta-phi}) is aligned with the gradient of the energy ($\nabla U$,EQ \ref{eq:energy}).

Another way of saying that the gradient of the energy and the gradient of the kinematics must point is the same direction, is that the angle between them is zero, or, using the dot product

\begin{equation}
    \nabla U \cdot \nabla f = |\nabla U| |\nabla f| \cos(0)
\end{equation}

This leaves three possibilities to satisfy this equation
\begin{enumerate}
    \item $\nabla U = 0$
    \item $\nabla f = 0$
    \item $\nabla f\cdot \nabla U = |\nabla U| |\nabla f|$
\end{enumerate}

%% file: supp/appendicies/energetics/combined-gradient.tex
In this section, we will show that the condition 

\begin{equation}
    \nabla f\cdot \nabla U = |\nabla U| |\nabla f|
\end{equation}

is the same as $\varphi = \pi$. Taking the gradients of the energy and kinematic level curve, we get that

\begin{equation}
    \nabla U = [ k_E(x-x_0), k_T(\theta-\theta_0), 0 ]
    \label{eq:nabla-U}
\end{equation}

and 

\begin{equation}
    \nabla f = \left[ \frac{\partial f}{\partial x}, \frac{\partial f}{\partial \theta}, \frac{\partial f}{\partial \varphi}\right]
\end{equation}

This implies that $\nabla U$ is always a 2D vector, whereas $\nabla f$ is always a 3D vector. The only possible way for these vectors to be pointing in the same direction is if 

\begin{equation}
    \frac{\partial f}{\partial \varphi} = 0
    \label{eq:partial-f-condition}
\end{equation}

If we take the partial derivative of EQ \ref{eq:f-x-theta-phi}, we see that 

\begin{equation}
    \frac{\partial f}{\partial \varphi} = - 2 \frac{\cos(\varphi/2)}{\sin\left(\eta + \sin^{-1}\left[\frac{x}{2\sin(\theta/2)}\right]\right)} = 0
\end{equation}

which can only be true if $\varphi = \pi$.

If we look at EQ \ref{eq:phi-2}, we see that in the special case of $\varphi_1 = \pi$, $\varphi_1 = \varphi_2$, so the two solutions collapse to one.

%% file: supp/appendicies/energetics/kinematic-gradient.tex
In this section we will show that $\nabla f \neq 0$ if $0<\eta <\pi$

The kinematic condition $\nabla f = 0$ expands to

\begin{equation}
    \nabla f = \left[ \frac{\partial f}{\partial x}, \frac{\partial f}{\partial \theta}, \frac{\partial f}{\partial \varphi}\right] = [0,0,0]
\end{equation}

Taking the $\theta$ term, we see that

\begin{equation}
    \frac{\partial f}{\partial \theta} = 0
    \label{eq:partial-theta-condition}
\end{equation}

We can rewrite EQ \ref{eq:f-x-theta-phi} as

\begin{equation}
    f(x,\theta,\varphi) = 0 = 2\pi \pm 2\sin^{-1}(\csc\gamma\sin(\theta/2)\sin(\eta+g(x,\theta))) - \varphi
\end{equation}

Taking the derivative, we get

\begin{equation}
    \frac{\partial f}{\partial \theta} = \frac{2\left(\csc\gamma\sin(\theta/2)\frac{\partial g}{\partial \theta}\cos(g +\eta) + \frac{1}{2}\csc\gamma\cos(\theta/2)\sin(g+\eta))\right)}{\sqrt{1 - \csc^2\gamma\sin^2(\theta/2)\sin^2(g+\eta)}}
\end{equation}

This can only be zero if 

\begin{equation}
    2\left(\csc\gamma\sin(\theta/2)\frac{\partial g}{\partial \theta}\cos(g +\eta) + \frac{1}{2}\csc\gamma\cos(\theta/2)\sin(g+\eta))\right) = 0
    \label{eq:zero-term-d-f-d-theta}
\end{equation}

Taking the derivative of EQ \ref{eq:g-of-x-and-theta}, we get

\begin{equation}
    \frac{\partial g}{\partial \theta} = \frac{x\cot(\theta/2)\csc(\theta/2)}{4\sqrt{1 - \frac{1}{4}x^2\csc^2(\theta/2)}} = \frac{1}{2}\tan g \cot(\theta/2)
\end{equation}

Substituting this into EQ \ref{eq:zero-term-d-f-d-theta}, and simplifying, we get

\begin{equation}
    \tan g = \tan(g+\eta)
\end{equation}

which is only true if $\eta = 0 \pm n\pi$. Because we have enforced, in Appendix \ref{section:parms-DOF}, that $0<\eta <\pi/2$, this will never be true, and therefore EQ \ref{eq:partial-theta-condition} will never be valid, and so $\nabla f \neq 0$.

%% file: supp/appendicies/energetics/global.tex
Inspecting EQ \ref{eq:nabla-U}, we see that $x = x_0$ and $\theta = \theta_0$ enforce that $\nabla U = 0$. As long as $x = x_0$ and $\theta = \theta_0$  are kinematically admissible, we know from Appendix \ref{section:x-theta-symmetry} that this corresponds to two solutions in $\alpha-\varphi$ space. These can be determined by plugging $x = x_0$ and $\theta = \theta_0$ into EQ's \ref{eq:phi-1} and \ref{eq:alpha-1} (for $\varphi_1$ and $\alpha_1$), and \ref{eq:phi-2} and \ref{eq:alpha-2} (for $\varphi_2$ and $\alpha_2$ ).

%% file: supp/appendicies/energetics/ke_kt_dependence.tex
\begin{figure}
    \centering
    \includegraphics{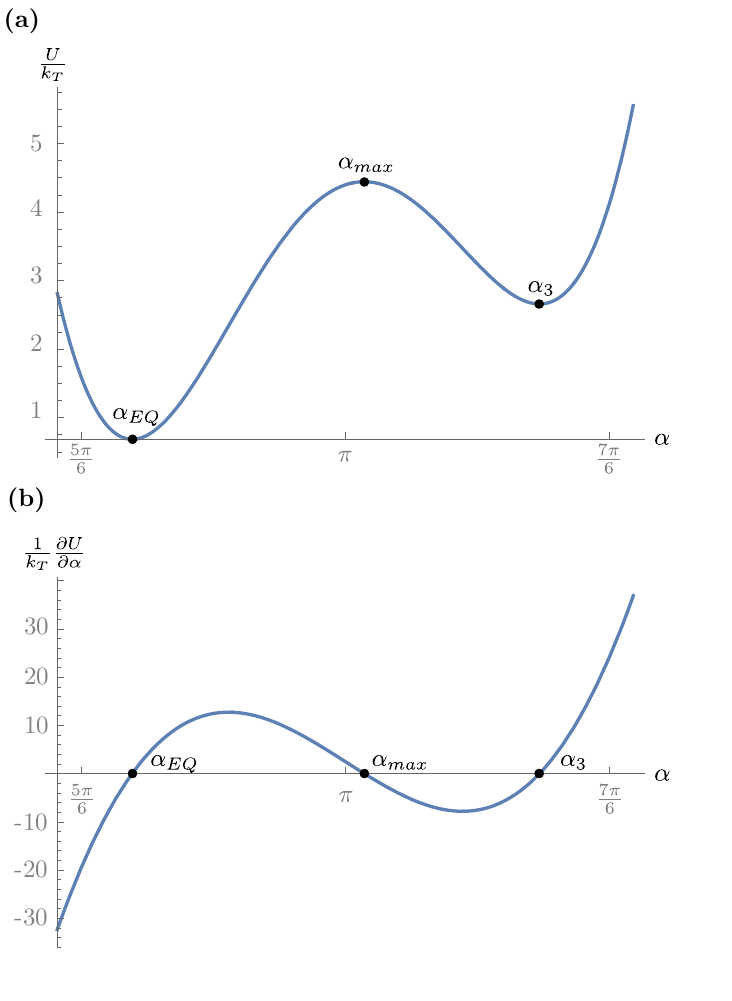}
    \caption{The local minima for this configuration}
    \label{fig:minima}
\end{figure}

In Appendix \ref{section:pi-equiv}, we showed that the local minimum will always be at $\varphi=\pi$. Since the system has two degrees of freedom, $\varphi$ and $\alpha$, and since one of the degrees of freedom is fixed at $\pi$, this simplifies to a one DOF optimization problem, which is fairly simple.

First, we rearrange EQ \ref{eq:energy-reproduced} to normalize it as follows

\begin{equation}
    U/k_T = \frac{1}{2} \left(\frac{k_E}{k_T}(x(\alpha, \pi)-x_0)^2 + (\theta(\alpha,\pi)-\theta_0)^2\right)
    \label{eq:energy-normalized}
\end{equation}

Figure \ref{fig:minima}a shows the energy around the equilibrium states (beyond this, it just blows up to large numbers). To produce this figure, we first plug in EQ\ref{eq:x-of-alpha-and-phi} and EQ\ref{eq:theta-of-alpha-and-phi} into EQ\ref{eq:energy-normalized} at $\varphi=\pi$. We then also plug the following parameters: $\gamma = 3\pi/4, \beta = 10 ^{\circ}, x_0 = 1/4,$ and $ \theta_0 = 1.39211$; into EQ\ref{eq:energy-normalized}. We also choose a ratio of $k_E/k_T = 10^3$ as discussed in the main text.

The local minimua occur when the slope of the function is zero, or

\begin{equation}
    \frac{\partial U}{\partial \alpha} = 0
\end{equation}

Which is equivalent to 

\begin{equation}
    \frac{1}{k_T}\frac{\partial U}{\partial \alpha} = 0
\end{equation}

We can find this, by taking the derivative of EQ\ref{eq:energy-normalized} with respect the $\alpha$, as follows

\begin{equation}
    \frac{1}{k_T}\frac{\partial U}{\partial \alpha} = \frac{k_E}{k_T}(x(\alpha, \pi)-x_0)\frac{\partial x(\alpha,\pi)}{\partial\alpha} + (\theta(\alpha,\pi)-\theta_0)\frac{\partial \theta(\alpha,\pi)}{\partial\alpha} = 0
    \label{eq:optimize-equation}
\end{equation}

The function \texttt{FindRoot[]} in Mathetmatica 12.3.1 yields three solutions for this equation, $\alpha_{EQ}, \alpha_{max},$ and $\alpha_3$, shown in the table below

\begin{table}[h]
    \centering
    \begin{tabular}{rl}
    \hline\hline
        State & $\alpha$ \\
        \hline
        EQ & 2.7192\\
        max & 3.1790\\
        T3 & 3.5267\\
        \hline\hline
    \end{tabular}
    \caption{The three roots of EQ \ref{eq:optimize-equation}}
    \label{tab:roots}
\end{table}

Notice that $\alpha_{max}$ is a maximum, so we can discard it. $\alpha_{EQ}$ turns out to be a saddle point, and is where the $\varphi$-path and the $\alpha$-path meet. It is a maximum on the $\varphi$-path, but a minimum on the $\alpha$-path. $\alpha_3$ is the only real local minimum we can find on this pathway. However, there is no mathematical guarantee that there will be only be one. What we can say, is that if there are multiple local minimum, they will have to exist along $\varphi=\pi$, and they will still be roots of EQ\ref{eq:optimize-equation}.

%% file: supp/appendicies/energetics/sans_stretch.tex
In this section, we will derive the equilibrium states for the non-Euclidean system without crease stretching. We begin by writing the whole energy of the system

\begin{equation}
    U = \frac{1}{2}k_T(\theta - \theta_0)^2
    \label{eq:intersect-energy}
\end{equation}

where $\theta_0 = 79.76^{\circ}$ as discussed in the main text, $\theta$ is defined in the same way as for the crease stretching system, $U$ is the energy, and $k_T$ is the torsional stiffness of the spring. Since we lie on the intersect boundary, and the intersect boundary (EQ \ref{eq:interset-dfn}) is only a function of $\alpha$, we can rewrite EQ \ref{eq:intersect-energy} as

\begin{equation}
    U(\alpha) = \frac{1}{2}k_T(\theta(\alpha) - \theta_0)^2
\end{equation}

and thus the stable states occur when 

\begin{equation}
    \frac{\partial U}{\partial \alpha} = 0 = k_T(\theta(\alpha) - \theta_0)\frac{\partial \theta}{\partial \alpha}
\end{equation}

which occurs either when 

\begin{equation}
    \frac{\partial \theta}{\partial \alpha} = 0 \quad \mathrm{or} \quad \theta(\alpha) = \theta_0
    \label{eq:choice}
\end{equation}

\subsubsection{$\partial \theta/\partial \alpha\neq 0$}
We begin by showing $\partial \theta/\partial \alpha$ is never zero. First, we we will emphasize values of $\theta$ which we are excluding. From Appendix \ref{section:closed-region}, we know that the $\alpha-\varphi$ pairs specified in Table \ref{tab:excluded-values} are part of the open boundary of the kinematic space and are excluded from analysis, because they are ambiguous (for instance $(\alpha,\varphi) = (0,0)$ is the same configuration as $(2\pi,2\pi)$)

\begin{table}[h]
    \centering
    \begin{tabular}{ccc}
    \hline\hline
        $\alpha$ & $\varphi$ & $\theta(\alpha,\varphi)$ \\
        \hline
        0 & 0 & 0\\
        $\pi$ & 0 & $2\pi$\\
        $\pi$ &$2\pi$ & $2\pi$\\
        $2\pi$ & $2\pi$ & 0\\
        \hline\hline
    \end{tabular}
    \caption{Points Excluded from Analysis Because They are Ambiguous and Their Corresponding $\theta$ Values}
    \label{tab:excluded-values}
\end{table}

We also calculate the corresponding $\theta$ values for these excluded points. They turn out to be 0 and $2\pi$, which implies the range of allowed $\theta$ are 

\begin{equation}
    \theta\in(0,2\pi)
\end{equation}

Now, using the four-part cotangent formula, we can relate $\theta$, $a$, $\alpha$, and $\gamma$ on triangle FAB in Figure \ref{fig:spheres-top} as

\begin{equation}
    \cos a\cos\alpha = \cot\gamma\sin a - \cot(\pi-\theta/2)\sin\alpha
    \label{eq:intersect-energy-begin}
\end{equation}

When there is no stretching involved, and C and D become the same point, we get that $a = \eta$. So Eq. \ref{eq:intersect-energy-begin} becomes 

\begin{equation}
    \cos\eta\cos\alpha = \cot\gamma\sin\eta + \cot(\theta/2)\sin\alpha
    \label{eq:no-stretching-cotangent-base}
\end{equation}

Then, taking the derivative with respect to $\alpha$ of both sides, we obtain

\begin{equation}
   -\cos\eta\sin\alpha = \cos\alpha\cot(\theta/2) - \frac{1}{2}\csc^2(\theta/2)\sin\alpha\frac{\partial\theta}{\partial\alpha}
\end{equation}

rearranging in terms of $\partial\theta/\partial\alpha$, we get

\begin{equation}
    \frac{\partial\theta}{\partial\alpha} = 2(\cos\eta + \cot\alpha\cot(\theta/2))\sin^2(\theta/2)
\end{equation}

since we know $\theta\in(0,2\pi)$, we know $\sin^2(\theta/2)$ will never be zero. Now, we must show that 

\begin{equation}
    \cos\eta +\cot\alpha\cot(\theta/2) \neq 0
\end{equation}

We will achieve this through proof by contradiction. So, we assume that

\begin{equation}
    \cos\eta +\cot\alpha\cot(\theta/2) = 0
    \label{eq:contradiction}
\end{equation}

which implies 

\begin{equation}
    \cot(\theta/2) = -\cos\eta\tan\alpha
\end{equation}

plugging this into Eq. \ref{eq:no-stretching-cotangent-base}, we get

\begin{equation}
    \cos\eta\cos\alpha = \cot\gamma\sin\eta - \cos\eta\tan\alpha\sin\alpha
\end{equation}

dividing by $\cos\eta$, we get 

\begin{equation}
    \cos\alpha = \cot\gamma\tan\eta - \frac{\sin^2\alpha}{\cos\alpha}
\end{equation}

multiplying both sides by $\cos\alpha$ and rearranging, we get

\begin{equation}
    \cos^2\alpha + \sin^2\alpha = \cot\gamma\tan\eta\cos\alpha
\end{equation}

or that

\begin{equation}
    1 = \cot\gamma\tan\eta\cos\alpha
\end{equation}

or that 

\begin{equation}
    \cos\alpha = \tan\gamma\cot\eta
\end{equation}

since $-1 \leq \cos\alpha \leq 1$, if we show that

\begin{equation}
    \tan\gamma\cot\eta < -1 
    \label{eq:inequality-gamma-eta}
\end{equation}

that implies that $\partial\theta/\partial\alpha \neq 0$. Since $0 < \eta < \pi/2$, and thus $\tan\eta$ is always positive, and never zero, we can rearrange Eq. \ref{eq:inequality-gamma-eta} as

\begin{equation}
    \tan\gamma < -\tan\eta
\end{equation}

substituting $\eta = \pi - \beta/2 -\gamma$ into this equation, we get 

\begin{equation}
    \tan\gamma < \tan(\beta/2 +\gamma)
    \label{eq:tangent-boys}
\end{equation}

We know that $\gamma$ and $\gamma+\beta/2$ always lie in the second quadrant, or more formally, reproducing Eq. \ref{eq:base-limits} and Eq. \ref{eq:beta-gamma-constraints}:

\begin{equation}
    \pi/2<\gamma < 3\pi/4\quad\mathrm{and}\quad\pi/2<\gamma + \beta/2 < \pi
\end{equation}

so Eq. \ref{eq:tangent-boys} becomes 

\begin{equation}
    \gamma < \beta/2 +\gamma
\end{equation}

which is always true if $\beta\neq0$. Thus Eq. \ref{eq:inequality-gamma-eta} is shown to be true, which falsifies Eq. \ref{eq:contradiction}, meaning

\begin{equation}
    \frac{\partial\theta}{\partial\alpha} \neq 0
\end{equation}

\subsubsection{$\alpha$ as a function of $\theta$}

Since $\partial\theta/\partial\alpha \neq 0$, according to Eq. \ref{eq:choice}, the only other place that stable states can occur is at

\begin{equation}
    \theta(\alpha) = \theta_0
\end{equation}

or, rewritten in a more useful way

\begin{equation}
    \alpha(\theta_0) = \alpha_{0,n}
\end{equation}

Where $\alpha_{0,n}$ are all the solutions that satisfy $\theta(\alpha) = \theta_0$. To find $\alpha(\theta)$, we begin by rearranging Eq. \ref{eq:no-stretching-cotangent-base} as follows 

\begin{equation}
    \cos\eta\cos\alpha - \cot(\theta/2)\sin\alpha = \cot\gamma\sin\eta
    \label{eq:theta-setup}
\end{equation}

we can now use the harmonic addition identity, which states that 

\begin{equation}
    a\cos\alpha + b\sin\alpha = c\cos(\alpha + \delta)
    \label{eq:harmoic-identity}
\end{equation}

where 

\begin{equation}
    c = \mathrm{sgn}(a)\sqrt{a^2 +b^2} \quad \mathrm{and} \quad \delta = \arctan\left[-\frac{b}{a}\right]\quad\forall a\neq 0
\end{equation}

plugging in $a = \cos\eta$ (which is never zero, since $0<\eta<\pi/2$) and $b =-\cot(\theta/2)$, we get 

\begin{equation}
    c = \sqrt{\cos^2\eta +\cot^2(\theta/2)} \quad \mathrm{and} \quad \delta = \arctan\left[\frac{\cot(\theta/2)}{\cos\eta}\right]
    \label{eq:a-b-dfn}
\end{equation}

Combining Eq. \ref{eq:a-b-dfn}, Eq. \ref{eq:harmoic-identity} and Eq. \ref{eq:theta-setup}, we get that

\begin{equation}
    \cot\gamma\sin\eta = \sqrt{\cos^2\eta +\cot^2(\theta/2)}\cos\left(\alpha + \arctan\left[\frac{\cot(\theta/2)}{\cos\eta}\right]\right)
    \label{eq:whole-shebang}
\end{equation}

Solving for $\alpha$, and plugging in $\theta_0$ for $\theta$, we get

\begin{equation}
\boxed{
    \alpha_{0,1}(\theta_0) = \mathrm{arccos}\left[\frac{\cot\gamma\sin\eta}{\sqrt{\cos^2\eta + \cot^2(\theta_0/2)}}\right] - \arctan\left[\frac{\cot(\theta_0/2)}{\cos\eta}\right]
    }
\end{equation}

Notice that Eq. \ref{eq:whole-shebang} can be rewritten as 

\begin{equation}
    \cot\gamma\sin\eta = \sqrt{\cos^2\eta +\cot^2(\theta/2)}\cos\left(2\pi - \alpha - \arctan\left[\frac{\cot(\theta/2)}{\cos\eta}\right]\right)
\end{equation}

Such that another solution emerges

\begin{equation}
\boxed{
    \alpha_{0,2}(\theta_0) = 2\pi -\mathrm{arccos}\left[\frac{\cot\gamma\sin\eta}{\sqrt{\cos^2\eta + \cot^2(\theta_0/2)}}\right] - \arctan\left[\frac{\cot(\theta_0/2)}{\cos\eta}\right]
    }
\end{equation}

% \tan\gamma < 0$ because as in Eq. , . We also know that $\tan(\gamma+\beta/2) < 0$ because, as in Eq. \ref{eq:beta-gamma-constraints}, $\pi/2<\gamma + \beta/2 < \pi$. Since $|\gamma| > |\beta/2 + \gamma|$ as long as $\beta\neq0$, and both tangents lie in the same quadrant and are always negative

% which yields an expression for $\partial\theta/\partial\alpha$:

% \begin{equation}
%     \frac{\partial\theta}{\partial\alpha} = \frac{2 \csc\alpha\sin(\theta/2)}{\cos\eta\sin(\theta/2) + \cot\alpha\cos(\theta/2)}
% \end{equation}

%% file: supp/appendicies/folding-paths/general.tex
\begin{figure}[h!]
    \centering
    \includegraphics{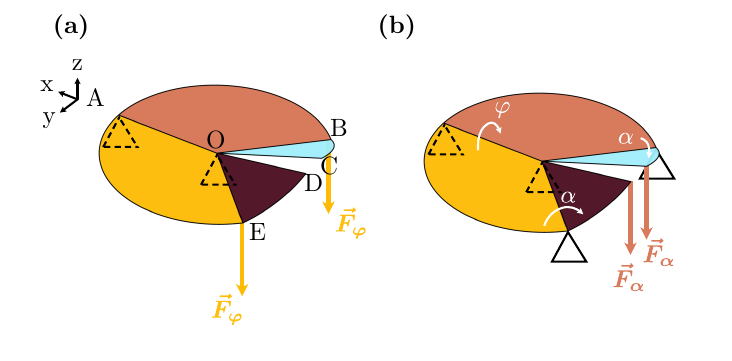}
    \caption{Applied Forces. In one pathway, we set $F_{\alpha} = 0$, in another we set $F_{\varphi} = 0$}
    \label{fig:DOF_with_forces}
    % figure updates:
    % - the triangle should be facing head on
\end{figure}

\subsection{Pathway Definitions}

We have found two pathways to move between stable states, the $\varphi$-path, where $\varphi$ is our degree of freedom, and the $\alpha$-path where $\alpha$ is our degree of freedom.

\subsubsection{$\varphi$-path Definition}\label{section:phi-path-dfn}

The $\varphi$-path is defined by applying two forces of magnitude $F_{\varphi}$ in the z-direction to points B and E, and keeping points O and A fixed. Written out mathematically, this is

\begin{equation}
    \vec{F}_B = \vec{F}_E = -F_{\varphi}(\varphi) \hat{k}
    \label{eq:phi-applied-force}
\end{equation}

with

\begin{equation}
    \vec{F}_C = \vec{F}_D = 0
    \label{eq:phi-zero-force}
\end{equation}

and 

\begin{equation}
    \frac{\partial\vec{r}_A}{\partial q_i} = \frac{\partial\vec{r}_O}{\partial q_i} = 0
\end{equation}

where $q_1$ is $\alpha$ and $q_2$ is $\varphi$, the two degrees of freedom. It is also easy to see by inspection that points B and E are both functions of $\varphi$ only, or

\begin{equation}
    \frac{\partial\vec{r}_E}{\partial \alpha} = \frac{\partial\vec{r}_B}{\partial \alpha} =  0
\end{equation}

Finally, we note that B and E are symmetric to each other about the x-z plane. In other words

\begin{equation}
    \vec{r}_E = X_{\varphi}(\varphi)\hat{i} + Y_{\varphi}(\varphi)\hat{j} + Z_{\varphi}(\varphi)\hat{k}
    \label{eq:re-dfn}
\end{equation}

\begin{equation}
    \vec{r}_B = X_{\varphi}(\varphi)\hat{i} - Y_{\varphi}(\varphi)\hat{j} + Z_{\varphi}(\varphi)\hat{k}
    \label{eq:rb-dfn}
\end{equation}

\subsubsection{$\alpha$-path Definition}\label{section:alpha-path-dfn}

The $\alpha$-path is defined by applying two forces of magnitude $F_{\alpha}$ in the z-direction to points $C$ and $D$, while keeping points A, O, E, and B fixed. Mathematically, this can be written as 

\begin{equation}
    \vec{F}_C = \vec{F}_D = -F_{\alpha}(\alpha)\hat{k}
    \label{eq:alpha-applied-force}
\end{equation}

and

\begin{equation}
    \frac{\partial\vec{r}_A}{\partial q_i} = \frac{\partial\vec{r}_B}{\partial q_i} = \frac{\partial\vec{r}_E}{\partial q_i} =\frac{\partial\vec{r}_O}{\partial q_i} = 0
    \label{eq:alpha-fixed points}
\end{equation}

since these conditions imply that $\varphi$ is fixed, we have that $\vec{r}_C$ and $\vec{r}_D$ are functions of $\alpha$ only, or that 

\begin{equation}
    \frac{\partial\vec{r}_C}{\partial \varphi} = \frac{\partial\vec{r}_D}{\partial \varphi} = 0
    \label{eq:c-d-phi-independence}
\end{equation}

Finally, we note that, as above, the vectors $\vec{r}_C$ and $\vec{r}_D$ are symmetric about the x-z plane, or that

\begin{equation}
    \vec{r}_D = X_{\alpha}(\alpha)\hat{i} + Y_{\alpha}(\alpha)\hat{j} + Z_{\alpha}(\alpha)\hat{k}
    \label{eq:rd-dfn}
\end{equation}

\begin{equation}
    \vec{r}_C = X_{\alpha}(\alpha)\hat{i} - Y_{\alpha}(\alpha)\hat{j} + Z_{\alpha}(\alpha)\hat{k}
    \label{eq:rc-dfn}
\end{equation}

\subsection{Principle of Virtual Work}

\subsubsection{Generalized Forces}

The principle of virtual work states that, for a system without applied moments

\begin{equation}
    \delta W = \sum_{k=1}^{f}\left[\left(\sum_{i=1}^{m}\vec{F_i}\cdot\frac{\partial \vec{r_i}}{\partial q_k}\right)\delta q_k \right] = \sum_{k=1}^{f}Q_k\delta q_k
\end{equation}

where $m$ is the number of points, $f$ is the number of generalized coordinated, $\vec{F}_i$ it the applied force at point i, $\vec{r}_i$ is the vector that points the point i, $q_k$ are the generalized coordinates (in this case, $q_1 = \alpha, q_2 =\varphi$) and $Q_k$ is the generalized force.

For this particular system, the virtual work looks like

\begin{multline}
\sum_{k=1}^{2}\Bigg[\bigg(
    \vec{F}_A\cdot\frac{\partial \vec{r}_A}{\partial q_k}
    +
    \vec{F}_B\cdot\frac{\partial \vec{r}_B}{\partial q_k}
    +
    \vec{F}_C\cdot\frac{\partial \vec{r}_C}{\partial q_k}
    +
    \vec{F}_D\cdot\frac{\partial \vec{r}_D}{\partial q_k}
    +
    \vec{F}_E\cdot\frac{\partial \vec{r}_E}{\partial q_k}
    +\\
    \vec{F}_O\cdot\frac{\partial \vec{r}_O}{\partial q_k}
    \bigg)\delta q_k\Bigg] = Q_{\varphi}\delta\varphi + Q_{\alpha}\delta\alpha
    \label{eq:expanded-lagrangian}
\end{multline}

Points A and O fixed in both pathways, so it follows that 

\begin{equation}
    \frac{\partial \vec{r}_A}{\partial q_k } = \frac{\partial \vec{r}_O}{\partial q_k } = 0
\end{equation}

so EQ\ref{eq:expanded-lagrangian} becomes

\begin{multline}
\sum_{k=1}^{2}\Bigg[\bigg(
    \vec{F}_B\cdot\frac{\partial \vec{r}_C}{\partial q_k}
    +
    \vec{F}_C\cdot\frac{\partial \vec{r}_C}{\partial q_k}
    +
    \vec{F}_D\cdot\frac{\partial \vec{r}_D}{\partial q_k}
    +
    \vec{F}_E\cdot\frac{\partial \vec{r}_E}{\partial q_k}
    \bigg)\delta q_k\Bigg] = \\Q_{\varphi}\delta\varphi + Q_{\alpha}\delta\alpha
\end{multline}

Expanded out, this becomes

\begin{multline}
\bigg(
    \vec{F}_B\cdot\frac{\partial \vec{r}_B}{\partial \varphi}
    +
    \vec{F}_C\cdot\frac{\partial \vec{r}_C}{\partial \varphi}
    +
    \vec{F}_D\cdot\frac{\partial \vec{r}_D}{\partial \varphi}
    +
    \vec{F}_E\cdot\frac{\partial \vec{r}_E}{\partial \varphi}
    \bigg)\delta \varphi 
    +\\
    \bigg(
    \vec{F}_B\cdot\frac{\partial \vec{r}_B}{\partial \alpha}
    +
    \vec{F}_C\cdot\frac{\partial \vec{r}_C}{\partial \alpha}
    +
    \vec{F}_D\cdot\frac{\partial \vec{r}_D}{\partial \alpha}
    +
    \vec{F}_E\cdot\frac{\partial \vec{r}_E}{\partial \alpha}
    \bigg)\delta \alpha 
    = \\Q_{\varphi}\delta\varphi + Q_{\alpha}\delta\alpha
\end{multline}

Since in all pathways $\vec{r}_E$ and $\vec{r}_B$ are not a function of $\varphi$, the virtual work reduces to 

\begin{multline}
\bigg(
    \vec{F}_B\cdot\frac{\partial \vec{r}_B}{\partial \varphi}
    +
    \vec{F}_C\cdot\frac{\partial \vec{r}_C}{\partial \varphi}
    +
    \vec{F}_D\cdot\frac{\partial \vec{r}_D}{\partial \varphi}
    +
    \vec{F}_E\cdot\frac{\partial \vec{r}_E}{\partial \varphi}
    \bigg)\delta \varphi 
    +\\
    \bigg(
    \vec{F}_C\cdot\frac{\partial \vec{r}_C}{\partial \alpha}
    +
    \vec{F}_D\cdot\frac{\partial \vec{r}_D}{\partial \alpha}
    \bigg)\delta \alpha 
    = Q_{\varphi}\delta\varphi + Q_{\alpha}\delta\alpha
    \label{eq:general-virtual-work}
\end{multline}

\subsubsection{The $\varphi$-pathway implies $\frac{\partial U}{\partial \alpha} = 0$}

We now simplify the virtual work for the $\varphi$-pathway. Inserting EQ\ref{eq:phi-applied-force} and EQ\ref{eq:phi-zero-force} into EQ\ref{eq:general-virtual-work}, we get

\begin{equation}
\bigg(
    -F_{\varphi} (\varphi)\hat{k}\cdot\frac{\partial \vec{r}_B}{\partial \varphi} + 
     -F_{\varphi} (\varphi)\hat{k}\cdot\frac{\partial \vec{r}_E}{\partial \varphi}
    \bigg)\delta \varphi 
    = Q_{\varphi}\delta\varphi + Q_{\alpha}\delta\alpha
\end{equation}

Inserting EQ\ref{eq:re-dfn} and EQ\ref{eq:rb-dfn}, we find that

\begin{equation}
    -2F_{\varphi}(\varphi)Z'_{\varphi}(\varphi)\delta \varphi 
    = Q_{\varphi}\delta\varphi + Q_{\alpha}\delta\alpha
\end{equation}

Because of the arbitrariness of the virtual displacements, we can conclude that

\begin{equation}
    Q_{\alpha}=0
\end{equation}

The generalized forces can be related to the energy potential of the system by

\begin{equation}
    Q_j = \frac{d}{dt}\frac{\partial U}{\partial \dot{q}_j} - \frac{\partial U}{\partial q_j}
\end{equation}

where $U$ is the energy potential from EQ\ref{eq:energy-reproduced}. If we assume quasi-static loading of the system, we have that 

\begin{equation}
    Q_j = - \frac{\partial U}{\partial q_j}
\end{equation}

Thus, we can conclude that the conditions for the $\varphi$-pathway in section \ref{section:phi-path-dfn} imply that $\partial U/\partial\alpha = 0$

\subsubsection{The $\alpha$-pathway implies $\frac{\partial U}{\partial \varphi} = 0$}

We now simplify the virtual work for the $\alpha$-pathway. Inserting EQ\ref{eq:phi-applied-force} and EQ\ref{eq:phi-zero-force} into EQ\ref{eq:general-virtual-work}, we get

\begin{equation}
    \bigg(
    -F_{\alpha}(\alpha)\hat{k}\cdot\frac{\partial \vec{r}_C}{\partial \alpha}
    +
    -F_{\alpha}(\alpha)\hat{k}\cdot\frac{\partial \vec{r}_D}{\partial \alpha}
    \bigg)\delta \alpha 
    = Q_{\varphi}\delta\varphi + Q_{\alpha}\delta\alpha
\end{equation}

Inserting EQ\ref{eq:rc-dfn} and EQ\ref{eq:rd-dfn}, we find that

\begin{equation}
    -2F_{\alpha}(\alpha)Z'_{\alpha}(\alpha)\delta \alpha 
    = Q_{\varphi}\delta\varphi + Q_{\alpha}\delta\alpha
\end{equation}

Because of the arbitrariness of the virtual displacements, we can conclude that

\begin{equation}
    Q_{\varphi}=0
\end{equation}

As shown in the previous section,

\begin{equation}
    Q_j = - \frac{\partial U}{\partial q_j}
\end{equation}

Thus, we can conclude that the conditions for the $\alpha$-pathway in section \ref{section:alpha-path-dfn} imply that $\partial U/\partial\varphi = 0$